\documentclass[twoside,twocolumn,9pt]{article}
\usepackage{extsizes}
\usepackage[super,sort&compress,comma]{natbib} 
\usepackage[version=3]{mhchem}
\usepackage[left=1.5cm, right=1.5cm, top=1.785cm, bottom=2.0cm]{geometry}
\usepackage{balance}
\usepackage{mathptmx}
\usepackage{sectsty}
\usepackage{graphicx} 
\usepackage{lastpage}
\usepackage[format=plain,justification=justified,singlelinecheck=false,font={stretch=1.125,small,sf},labelfont=bf,labelsep=space]{caption}
\usepackage{float}
\usepackage{fancyhdr}
\usepackage{fnpos}
\usepackage[english]{babel}
\addto{\captionsenglish}{%
  \renewcommand{\refname}{Notes and references}
}
\usepackage{array}
\usepackage{droidsans}
\usepackage{charter}
\usepackage[T1]{fontenc}
\usepackage[usenames,dvipsnames]{xcolor}
\usepackage{setspace}
\usepackage[compact]{titlesec}
\usepackage{hyperref}
\usepackage{color}

\usepackage{epstopdf}

\definecolor{cream}{RGB}{222,217,201}

\begin{document}

\pagestyle{fancy}
\thispagestyle{plain}
\fancypagestyle{plain}{
\renewcommand{\headrulewidth}{0pt}
}

\makeFNbottom
\makeatletter
\renewcommand\LARGE{\@setfontsize\LARGE{15pt}{17}}
\renewcommand\Large{\@setfontsize\Large{12pt}{14}}
\renewcommand\large{\@setfontsize\large{10pt}{12}}
\renewcommand\footnotesize{\@setfontsize\footnotesize{7pt}{10}}
\makeatother

\renewcommand{\thefootnote}{\fnsymbol{footnote}}
\renewcommand\footnoterule{\vspace*{1pt}%
\color{cream}\hrule width 3.5in height 0.4pt \color{black}\vspace*{5pt}} 
\setcounter{secnumdepth}{5}

\makeatletter 
\renewcommand\@biblabel[1]{#1}            
\renewcommand\@makefntext[1]%
{\noindent\makebox[0pt][r]{\@thefnmark\,}#1}
\makeatother 
\renewcommand{\figurename}{\small{Fig.}~}
\sectionfont{\sffamily\Large}
\subsectionfont{\normalsize}
\subsubsectionfont{\bf}
\setstretch{1.125} 
\setlength{\skip\footins}{0.8cm}
\setlength{\footnotesep}{0.25cm}
\setlength{\jot}{10pt}
\titlespacing*{\section}{0pt}{4pt}{4pt}
\titlespacing*{\subsection}{0pt}{15pt}{1pt}

\fancyfoot{}
\fancyfoot[LO,RE]{\vspace{-7.1pt}\includegraphics[height=9pt]{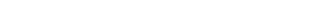}}
\fancyfoot[CO]{\vspace{-7.1pt}\hspace{13.2cm}\includegraphics{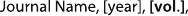}}
\fancyfoot[CE]{\vspace{-7.2pt}\hspace{-14.2cm}\includegraphics{head_foot/RF}}
\fancyfoot[RO]{\footnotesize{\sffamily{1--\pageref{LastPage} ~\textbar  \hspace{2pt}\thepage}}}
\fancyfoot[LE]{\footnotesize{\sffamily{\thepage~\textbar\hspace{3.45cm} 1--\pageref{LastPage}}}}
\fancyhead{}
\renewcommand{\headrulewidth}{0pt} 
\renewcommand{\footrulewidth}{0pt}
\setlength{\arrayrulewidth}{1pt}
\setlength{\columnsep}{6.5mm}
\setlength\bibsep{1pt}

\makeatletter 
\newlength{\figrulesep} 
\setlength{\figrulesep}{0.5\textfloatsep} 

\newcommand{\topfigrule}{\vspace*{-1pt}%
\noindent{\color{cream}\rule[-\figrulesep]{\columnwidth}{1.5pt}} }

\newcommand{\botfigrule}{\vspace*{-2pt}%
\noindent{\color{cream}\rule[\figrulesep]{\columnwidth}{1.5pt}} }

\newcommand{\dblfigrule}{\vspace*{-1pt}%
\noindent{\color{cream}\rule[-\figrulesep]{\textwidth}{1.5pt}} }

\makeatother

\twocolumn[
  \begin{@twocolumnfalse}
{\includegraphics[height=30pt]{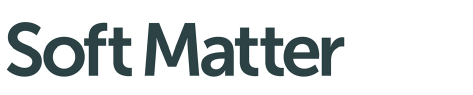}\hfill\raisebox{0pt}[0pt][0pt]{\includegraphics[height=55pt]{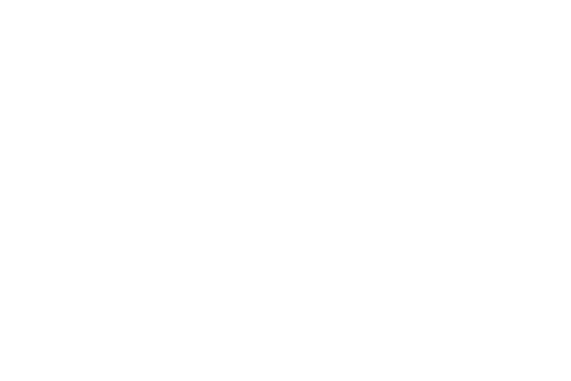}}\\[1ex]
\includegraphics[width=18.5cm]{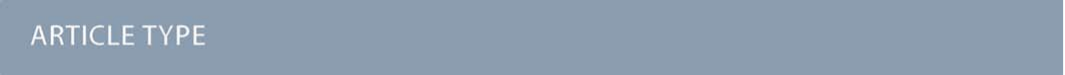}}\par
\vspace{1em}
\sffamily
\begin{tabular}{m{4.5cm} p{13.5cm} }

\includegraphics{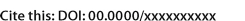} & \noindent\LARGE{\textbf{Spatiotemporal characterization of emergent behavior of self-propelled oil droplet$^\dag$}} \\
\vspace{0.3cm} & \vspace{0.3cm} \\

 & \noindent\large{Riku Adachi,\textit{$^{a}$} Hiroki Kojima,\textit{$^{a}$} and Takashi Ikegami\textit{$^{a}$}} \\

\includegraphics{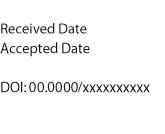} & \noindent\normalsize{To further understand the complex behavior of swimming microorganisms, the spontaneous motion of nonliving matter provides essential insights. While substantial research has focused on quantitatively analyzing complex behavioral patterns, characterizing these dynamics aiming for inclusive comparison to the behavior of living systems remains challenging. In this study, we experimentally and numerically investigated the 'life-like' behavior of an oil droplet in an aqueous surfactant solution by identifying behavioral modes of spontaneous motion patterns in response to varying physical parameters, such as volume and oil concentration. Leveraging data-driven nonparametric dynamical systems analysis, we discovered the low dimensionality and nonlinearity of the underlying dynamical system governing oil droplet motion. Notably, our simulations demonstrate that the two-dimensional Langevin equations effectively reproduce the overall behavior experimentally observed while retaining the rational correspondence with physical parameter interpretations. These findings not only elucidate the fundamental dynamics governing the spontaneous motion of oil droplet systems but also suggest potential pathways for developing more descriptive models that bridge the gap between nonliving and living behaviors.} \\

\end{tabular}

 \end{@twocolumnfalse} \vspace{0.6cm}

  ]

\renewcommand*\rmdefault{bch}\normalfont\upshape
\rmfamily
\section*{}
\vspace{-1cm}


\footnotetext{\textit{$^{a}$~Graduate School of Arts and Sciences, University of Tokyo, 3-8-1 Komaba, Meguro-ku 153-8902, Tokyo, Japan. E-mail: adachi@sacral.c.u-tokyo.ac.jp}}

\footnotetext{\dag~Supplementary Information available: [details of any supplementary information available should be included here]. See DOI:}




\section{Introduction}
Self-propelled particles (SPPs) are spatially and temporally ubiquitous across natural environments and laboratories \cite{Ramaswamy2010-ok, Shaebani2020-et}. Examples of SPPs span various scales and physical properties, ranging from collective motion in microorganisms and self-propelled active droplets to the cohesive movements observed in schools of fish \cite{Katz2011-au} and flocks of birds \cite{Ballerini2008-wt}. These examples encapsulate living and nonliving matters within the spectrum of active matter physics \cite{Bowick2022-ti, Marchetti2013-da}. Living systems exhibit a remarkable range of behaviors, from the intricate locomotion of microorganisms to the coordinated dynamics of animal swarms \cite{Tu2018-vu}. In contrast, nonliving systems, under specific conditions, can display behaviors that mimic life, offering a simplified yet illuminating framework for constructing the physics of behavior. A critical challenge lies in bridging these diverse domains—developing a unified phenomenology to quantitatively compare and analyze behaviors across living and nonliving systems. 

Among them, chemically active swimming droplets are an emerging realm in these decades, and research has been dedicated to understanding the mechanism of self-propulsion and explaining the behavior experimentally and theoretically \cite{Maass2016-wl, Hokmabad2021-px, Michelin2023-ut}. Insofar oil droplets in an aqueous surfactant solution or water droplets in the oil phase can perform spontaneous motion, has been reported \cite{Toyota2009-ge, Horibe2011-sy, Sumino2005-ep, Nagai2005-zy, Kruger2016-lq, Suda2022-yz}. Pioneering works by Hanczyc et al \cite{Hanczyc2007-pc}. discovered that fatty acid anhydride precursor introduced into an aqueous fatty acid micelle solution shows autonomous, sustained movements in the aqueous medium—conversely, Izri et al \cite{Izri2014-vc}. experimentally uncovered that the pure water droplets swim spontaneously in the oil (squalane)–surfactant (monoolein) medium. Interestingly, due to the nonlinear physicochemical coupling under viscous fluid environments, these chemically active SPPs exhibit even more complex 'life-like' behavior, including growth and division \cite{Hanczyc2003-ra, Zwicker2016-ic}, predator-prey interaction \cite{Meredith2020-ur, Liu2024-kt}, solving a maze \cite{Lagzi2010-ad, Jin2017-yn}, and collective behaviors \cite{Tanaka2017-vu, Jin2021-fn}. Although the term 'life-like behavior' has often been used to describe the complex motions of artificial SPPs, it remains loosely defined. In this study, we define 'life-like behavior' as the self-sustained, adaptive, and nonlinear dynamics exhibited by systems driven far from equilibrium. This definition offers a structured approach to quantitatively analyze the emergent behaviors of nonliving matter and their resemblance to biological systems.

The mechanistic origin driving these droplet systems is typically Marangoni flow from the inhomogeneous surface tension gradients \cite{Pimienta2014-fh}. Emanated products from a droplet form the concentration gradients of a surface active compound via the chemical reaction or micelle solubilization. Initially, a droplet dissolves uniformly, yet either the internal and external fluctuations or anisotropy of the droplet shape triggers the breaking of spatial symmetry, which leads droplets to initiate self-propulsion and consequentially climb up the surface tension gradient. Once the motion is initiated, the dissymmetry is maintained by the feedback loop created by restoring surface tension after their passing trajectories. Droplets consume chemical energy to generate mechanical work, and they can migrate within the out-of-equilibrium environment until relaxation in an equilibrium state. Additionally, this driving mechanism is coupled with chemical gradients in the surface produced due to a chemical reaction between the droplet and its environment, referred to as chemotaxis. They perceive the gradient concentration as input and subsequently generate the migration towards chemoattractants or away from chemorepellants as output motion  \cite{Hanczyc2010-jx}. The interplay between environmental gradients and surface tension-driven flows highlights the inherently nonlinear nature of a droplet's complex spontaneous behavior. The feedback mechanisms maintaining asymmetry underscore the complexity of these systems, as emergent patterns depend sensitively on initial conditions and external perturbations. Understanding this nonlinearity is critical for capturing the full spectrum of behaviors, particularly when comparing them to naturalistic dynamics in living systems \cite{suematsu2018evolution}. Although the behavior of living systems is characterized by nonlinearity, to our knowledge, previous analyses of self-propelled droplets have been limited to basic statistics such as velocity and velocity autocorrelations, and the nonlinearity within their 'life-like' behavior has not yet been quantitatively analyzed. Additionally, in an attempt to construct a quantitative phenomenology of naturalistic behavior, the dimensionality of the underlying dynamical system of behavior is a promising candidate \cite{Berman2016-ut, Bialek2022-pl}. For example, Ahamed et al \cite{Ahamed2020-np}. reported that the dynamics of the posture change of C.elegans can be reduced to a space of about seven dimensions. It is also known that the dynamics of a fruit fly can be explained to a great extent by the superposition of a small number of linear modes \cite{berman2014mapping}. Suppose the dimension of governing a dynamical system parametrizes a continuous spectrum of living organisms and nonliving matters; the dimensionality of self-propelled droplets may provide a comparison benchmark. By employing a data-driven framework, this study addresses the challenge of characterizing such nonlinearities and dimensionality and provides a foundation for bridging the behaviors of living and nonliving systems.

With the above considerations, we experimentally explore the organization of the spontaneous trajectory patterns of a binary mixture oil droplet while sweeping physical parameters such as volumes and composition ratios floated on the aqueous surfactant solution. We investigated systems of millimeter-sized ethyl salicylate (ES) and paraffin droplets in an aqueous surfactant solution of sodium dodecyl sulfate (SDS), initially reported by Tanaka et al \cite{Tanaka2015-sr} (Video S1, ESI$\dag$). This experimental setup is advantageous in terms of its simplicity, tunability, scalability, and long-lived properties compared to the micro-sized droplets exhibiting active Brownian motions, which are primarily focused on these days. The dissolution of ES from a droplet reduces the surface tension surrounding the droplet, initiating a chemical Marangoni flow away from it. This flow enhances the dissolution process and generates inhomogeneities in the dissolution. Motility can be induced in surface tension differences at the front and rear of the droplet, giving rise to translational (e.g., straight, circular, and knot-forming), even nonlinear motion (e.g., back-and-forth, and irregular) while dynamically deforming their shape \cite{Satoh2017-oz, Tanaka2021-ax}. The mode of motion is determined by whether the reaction force passes through the mass center of the droplet. The perpendicular force acting on the spherical droplet always passes through its center of mass, thus leading to translational motion. For the deformed anisotropic droplet, the reaction force does not pass the center of mass, and the generated torque induces helical or circular motion. This kind of rotational trajectory has been discovered in biological behavior, such as the swimming patterns of the Tetrahymena cells \cite{Marumo2021-ua} and E.coli \cite{Lauga2009-iz, Perez_Ipina2019-ot} with structural asymmetry both in experiments and simulations. We draw the phase diagram classifying spatiotemporal behavioral patterns and implement the quantitative evaluation employing data-driven dynamical system analysis, including embedding dimension estimation, nonlinearity analysis, and computation of the Lyapunov spectrum. Subsequently, to rationalize the description of nonlinear dynamical systems, we performed a minimal phenomenological simulation of the two-dimensional Langevin equations comprising driving forces to reproduce observed movement. By investigating the underlying dynamics of the spontaneous motion of artificial swimmers, we aim to pave the way for the quantitative comparison of the continuity of life and non-life through an abstract physics perspective on motility.  

\section{Experiment}
\subsection*{Experimental setup}
\begin{figure}[htbp]
\centering
  \includegraphics[keepaspectratio, width=\linewidth]{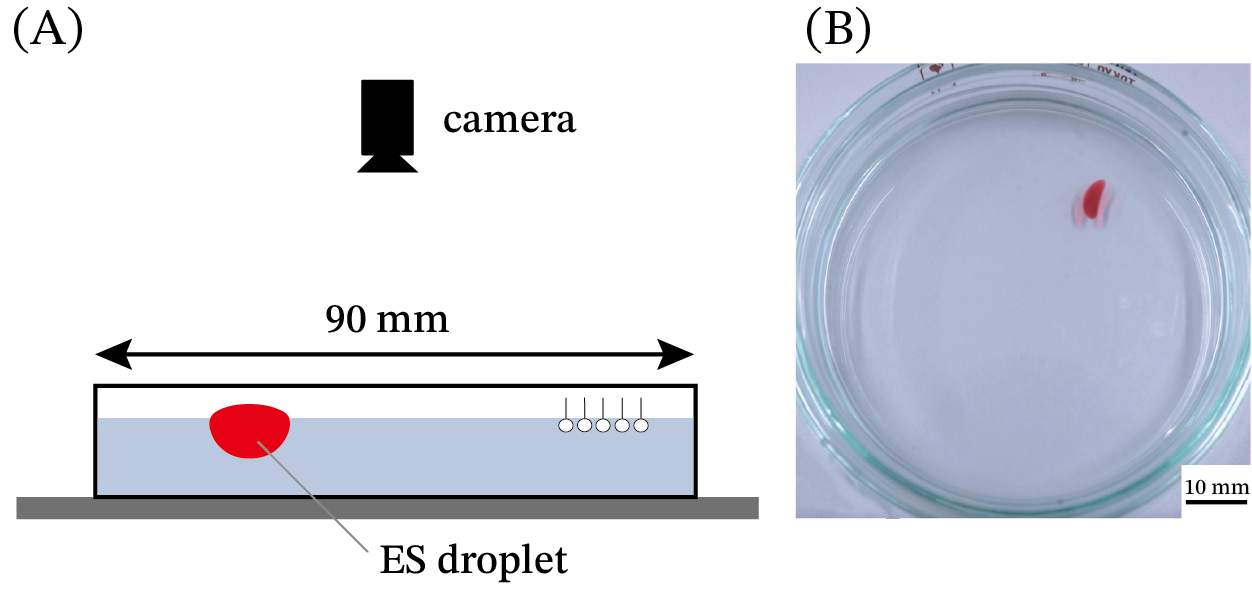}
  \caption{(A) Schematic illustration of the experimental setup. (B) The top view of the swimming droplet with a volume of 15 $\mathrm{\, \mu L}$ and 60 \% ES concentration. }
  \label{fig:fig1}
\end{figure}

A single oil droplet consisting of a mixture of ethyl salicylate (ES) and paraffin was placed in a petri dish containing an anionic surfactant, Sodium Dodecyl Sulfate (SDS) solution. By changing the key physical parameters: volume (5, 10, 15, 20, 25, and 30 $\mathrm{\, \mu L}$) and ES concentration  $c_{ES}$ (50, 60, 70, 80, and 90 (wt. \%)) of an oil droplet, the emerging motion was recorded. The Reynolds number $Re = UL/ \nu$ of this system is approximately on the order of $\sim 10^{2}$ with the typical length $L \sim 10^{-2} \mathrm{\, m}$, fluid velocity $U \sim 10^{-2} \mathrm{\, ms^{-1}}$), and kinematic viscosity of water $\nu \sim 10^{-2} \mathrm{\, m^2s^{-1}}$. Since the diffusion coefficient $D$ of a molecule in water is roughly $D \sim 10^{-9} \mathrm{\, m^2s^{-1}}$, the Schmidt number  $Sc$ is estimated to be $Sc := \nu / D \sim 10^{3}$. Therefore, the Péclet number $Pe$ is defined as $Pe := UL/D = Re \times Sc \sim 10^{5}$, meaning the advection is completely dominant over the diffusion. For the real-time recording of the motion of the oil droplet, we utilized a digital video camera (ILCE-7M3, SONY) mounted vertically above the arena (Fig.\ref{fig:fig1}. They were recorded 30 minutes long, at least five trials per one pair of parameter configurations. The frame rate of the video recording was 24 frames per second, and each image was 1920 $\times$ 1080 pixels. The time series of the centroid displacements of a droplet was extracted through the color filter using Python and Open-CV software. 

\subsection*{Data acquisition and image processing}
The minute amount of Oil red O was added as a dye for visualization. The concentration of the SDS solution remained constant at 69 $\mathrm{\, mM}$, well above the critical micelle concentration of approximately 8 $\mathrm{\, mM}$. Following the placement of a droplet onto a glass dish with a 90 $\mathrm{\, mm}$ inner diameter and containing 30 $\mathrm{\, mL}$ of SDS, the dish was covered with a glass lid without being tightly sealed. The room temperature was controlled at about 22 ${}^\circ$C by an air conditioner. 

\subsection*{Materials}
The ES was purchased from Tokyo Chemical Industry Co., Ltd., and the paraffin liquid was purchased from Sigma-Aldrich Japan. Both chemicals were used as supplied. Oil red O and SDS were purchased from Wako Pure Chemical Industries. Purified water for the SDS solution was purchased from AS ONE Corporation. 

\section{Results and discussion}
\subsection*{Spatiotemporal characterization and statistics of an oil droplet behavior}
\begin{figure*}[htbp]
 \centering
 \includegraphics[width = \linewidth, keepaspectratio]{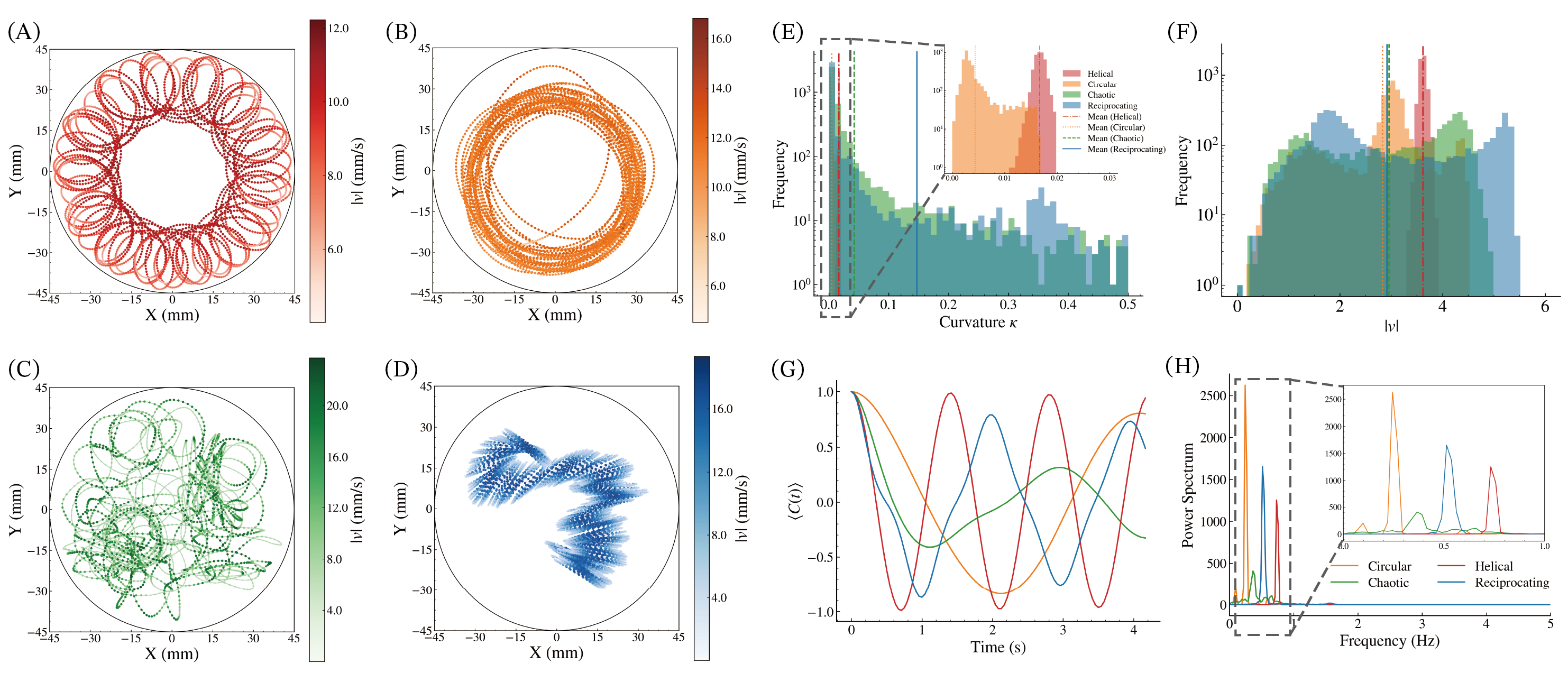}
 \caption{Characteristic trajectory patterns of the center of mass of an oil droplet and their spatial and temporal characterization. (A-D) Representative three-minute trajectories of the droplet's center of mass for rotational patterns—helical (large volume) and circular (small volume, high $c_{ES}$)—chaotic motion (mid volume, mid $c_{ES}$), and reciprocating motion (small volume, high $c_{ES}$). Each color bar indicates the magnitude of velocities $|v|:= \sqrt{v_x^2 + v_y^2}$, and the black circle represents the dish wall. (E) Spatial characterization of these patterns, including histograms of curvature with zoomed views (inset) and (F) velocity distributions. Colored lines indicate the mean of each distribution derived from segmented experiments under varying physical parameters. (G) Temporal characterization of these patterns, featuring the autocorrelation function of droplet velocities and their power spectra with detailed views (inset).}
 \label{fig:fig11}
\end{figure*}

Fig.\ref{fig:fig11} represents examples of oil droplet trajectories. The typical motion patterns can be classified into four categories: rotational (i.e., helical (Fig.\ref{fig:fig11} A) or circular (Fig.\ref{fig:fig11} B) depending on the amplitudes), chaotic (Fig.\ref{fig:fig11} C), reciprocating (Fig.\ref{fig:fig11} D), and stationary. From this point forward, we focus on the diverse motile behaviors only as they offer critical insights into the dynamic properties of motion. While this classification respects the qualitative framework of previous studies, note that they often rely on subjective parameter choices that reflect the researcher's perceptions of behavior rather than capturing its full diversity. To address this, we adopt the following model-independent, data-driven approach to classify and spatially and temporally characterize these motion dynamics of self-propelled droplets.

To classify and characterize the motion patterns of self-propelled oil droplets, we adopt a model-independent approach leveraging both spatial and temporal descriptors. Since the droplets dissolve gradually, neither speed nor diameter are constant over time, which has to be considered in long-lasting motion. Thus, we divide 30-minute trajectories into non-overlapping windows of length $t_w$ (3 min), and within each window, we obtain the $x$-directional velocity $v_x$ and $|v|$ distribution, which we analyze below. Spatial features are captured using curvature and velocity distribution: curvature quantifies the geometric properties of the trajectory, such as bending intensity, while the velocity distribution reflects the variability and magnitude of the droplet's movement, providing insights into the spatial dynamics of the system. Temporal features are assessed through the autocorrelation function (ACF) and power spectrum (PS): the autocorrelation function evaluates time-dependent persistence and periodicity in motion, while the power spectrum reveals dominant frequency components and the degree of motion regularity in the frequency domain. These descriptors are derived directly from the observed data without assuming any underlying models, ensuring an unbiased and data-driven framework for motion classification. Furthermore, the combination of spatial and temporal metrics allows for a comprehensive characterization of the system's spatiotemporal dynamics, enabling us to objectively distinguish between different motion patterns.

The spatial characteristics within behavioral patterns are quantified by the velocity distributions and by tracking the local curvature $\kappa (t)$ of the droplet's center of mass trajectory $\mathbf x_j = (x(t_j), y(t_j))$, sampled at times $t_j = j \cdot \Delta t$ for $j = 1, \cdots, N$ with  $\Delta t = f^{-1}_M$, the inverse of the recording frame rate:
$$
\kappa (t_i) = \Big \langle \frac{\dot x(t_i) \ddot y (t_i) - \dot y(t_i) \ddot x(t_i)}{(\dot x (t_i)^2 + \dot y (t_i)^2)^{3/2} } \Big\rangle_{t_w}
$$

The average $\langle \cdot \rangle$ is taken over the divided time window $t_w$. The local curvature measurements naturally reflect the geometric properties of the motion. For example, in rotational trajectories, differences in radius $R$ result in corresponding low variations in curvature, as expected from the relationship $\kappa = 1/R$. Thus, it enables curvatures to serve as robust descriptors of trajectory geometry across different motion patterns. As shown in Fig.\ref{fig:fig11} (E), when the trajectory forms rotations with a radius similar to the arena diameter and low local curvature (practically, $ 0.03 < \kappa$), the motion is helical or circular depending on the magnitude of amplitude. These rotational motions result from boundary-induced repulsion bending unidirectional motion. They were found to be achiral, with clockwise and counter-clockwise spirals observed with equal probability. Chaotic patterns, characterized by moderate curvature ($ 0.03 \le \kappa < 0.1$) and intermittent rather than periodic motions, show irregular movements and changes in direction without consistent displacement. The quasi-1D reciprocating motion was observed with a large curvature value ($ \kappa \ge 0.1$), characterized by periodic bidirectional patterns. The thresholds of each pattern were obtained as local minima of the curvature distribution by employing the Freedman–Diaconis rule.

Fig.\ref{fig:fig11} (F) presents the velocity distribution of corresponding motion patterns, respectively. Since droplets swim at a constant speed unidirectionally, the distributions become Gaussian-like unimodal shapes. The width of the distributions shows the intensity and amplitudes of rotational motions, which differentiate helical and circular. While chaotic and reciprocating patterns possess two motion phases, run-and-stop, the bimodal distributions were obtained.

The temporal characteristics within behavioral patterns are quantified by the power spectrum density of velocity $\mathbf v$ and  the velocity autocorrelation function of each pattern as a function of time $C(t)$ as : 
$$
C(t) = \Big \langle \frac{\mathbf v (t_0 + t) \cdot \mathbf v(t)}{ |\mathbf v (t_0 +t)| |\mathbf v(t_0)|} \Big \rangle_{t_w}
$$  
Fig.\ref{fig:fig11} (G) shows $C(t)$ obtained from four classes of motion patterns, respectively. The characteristic oscillation $\propto \cos (2 \pi T')$ with rotation period $T'$ varies from 1.5 $\text{s}$ to 4 $\text{s}$. The increase in the rotational period $T'$ appears from a lower droplet velocity, which in turn occurs because of a lower ES concentration around the surface of the droplet accompanied by a descending solubilization rate. The collapse of periodicity is observed within the chaotic behavior resulting from the intermittency of the motion. Sharp and strong peaks of the power spectrum corresponding to periodic patterns (i.e., helical, circular, and reciprocating) are shown in Fig.\ref{fig:fig11} (H), which imply that the space of possible behavioral patterns is limited in terms of the degrees of freedom and governed by the linear combination of these specific frequency modes $\sim 1 \mathrm{\, s^{-1}}$. In contrast, the power spectrum of the chaotic pattern exhibits the fluctuations around the primary oscillation mode $\sim 0.4 \mathrm{\, s^{-1}}$, indicative of the nonlinear behavioral dynamics within the behavioral state space. 

\begin{figure}[htbp]
\centering
  \includegraphics[keepaspectratio, scale = 0.75]{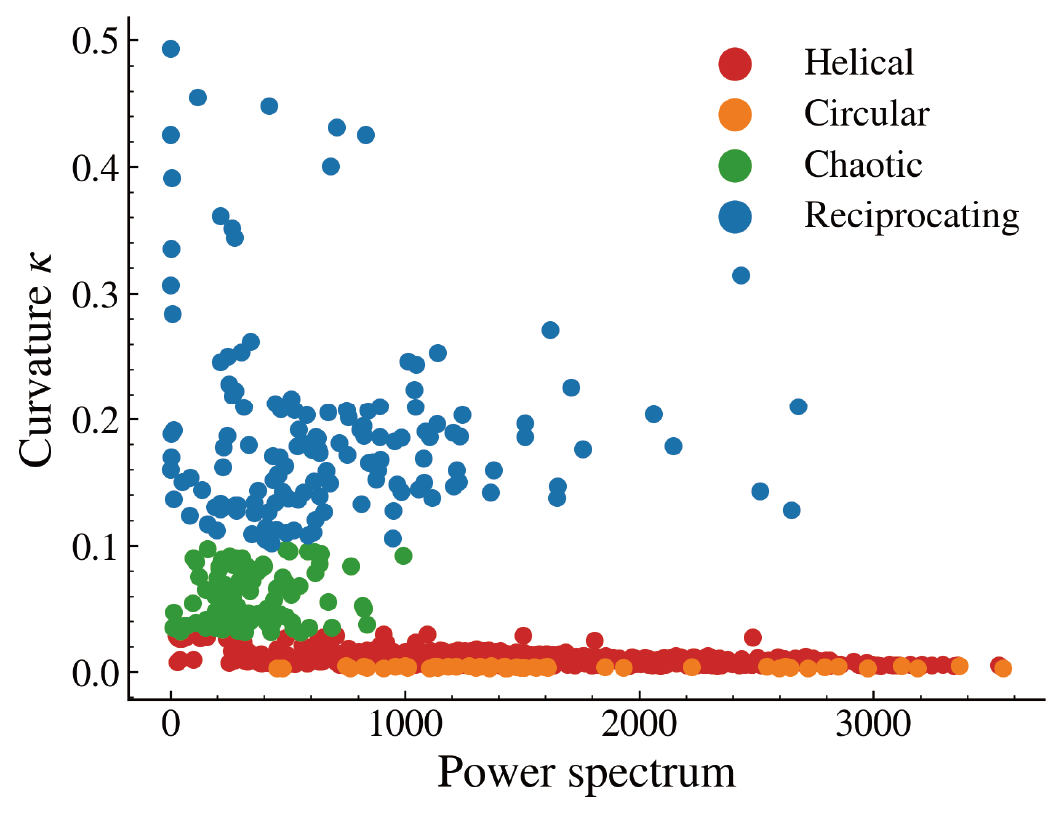}
  \caption{Feature space of the droplet behavior by the power spectrum peak intensity versus local curvature.}
  \label{fig:fig5}
\end{figure}

Overall, Table.\ref{tbl:table1} summarizes the criteria of spatiotemporal characterization and allows us the model-free quantitative classification of four types of behaviors - circular, helical, chaotic, and reciprocating. For example, in Fig.\ref{fig:fig5}, we plotted the peak intensity of the power spectrum versus local curvature across the time windows of all experimental trials. These behavioral patterns occupy different regions separately within a statistical feature space. But how do these behaviors correspond to the physical parameter configurations of the experiment? Also, how do these dynamical behaviors undergo transition with time evolution?

\begin{table*}
\centering 
\renewcommand{\arraystretch}{1.2} 
\small
  \caption{Summary of spatiotemporal characterizations and counts of segments classified as each pattern averaged over across all experiments.}
  \label{tbl:table1}
  \begin{tabular*}{\textwidth}{@{\extracolsep{\fill}}lllll}
    \hline
     & Circular & Helical & Chaotic & Reciprocating\\
    \hline
    Curvature  & $\kappa \le$ 0.005 & $0.005 < \kappa \le 0.03$ & $0.03 < \kappa \le 0.1 $ & $0.1 < \kappa$\\
    Vel. dist. & unimodal & unimodal & bimodal & bimodal \\
    PS freq ($\mu \pm \sigma$, $\mathrm{\, s^{-1}}$) & $0.18 \pm 0.05$ & $0.47 \pm 0.14$ & $0.34 \pm 0.11$  & $0.35 \pm 0.13$ \\
    PS peak ($\mu \pm \sigma$) & $1670 \pm 801$  & $1429 \pm 678$ & $348 \pm 193$ & $700 \pm 549$ \\
    ACF (characteristic time, $\mathrm{\, s}$) & 1.6, periodic & 0.6, periodic & 0.7, aperiodic & 0.8, periodic  \\
    \# of segments & 52 & 627 & 98 & 143\\ 
    \hline
  \end{tabular*}
\end{table*}

\subsection*{Dependence on physical parameters and spatial and temporal evolution of behavior}
\begin{figure}[htbp]
\centering
  \includegraphics[keepaspectratio, width=\linewidth]{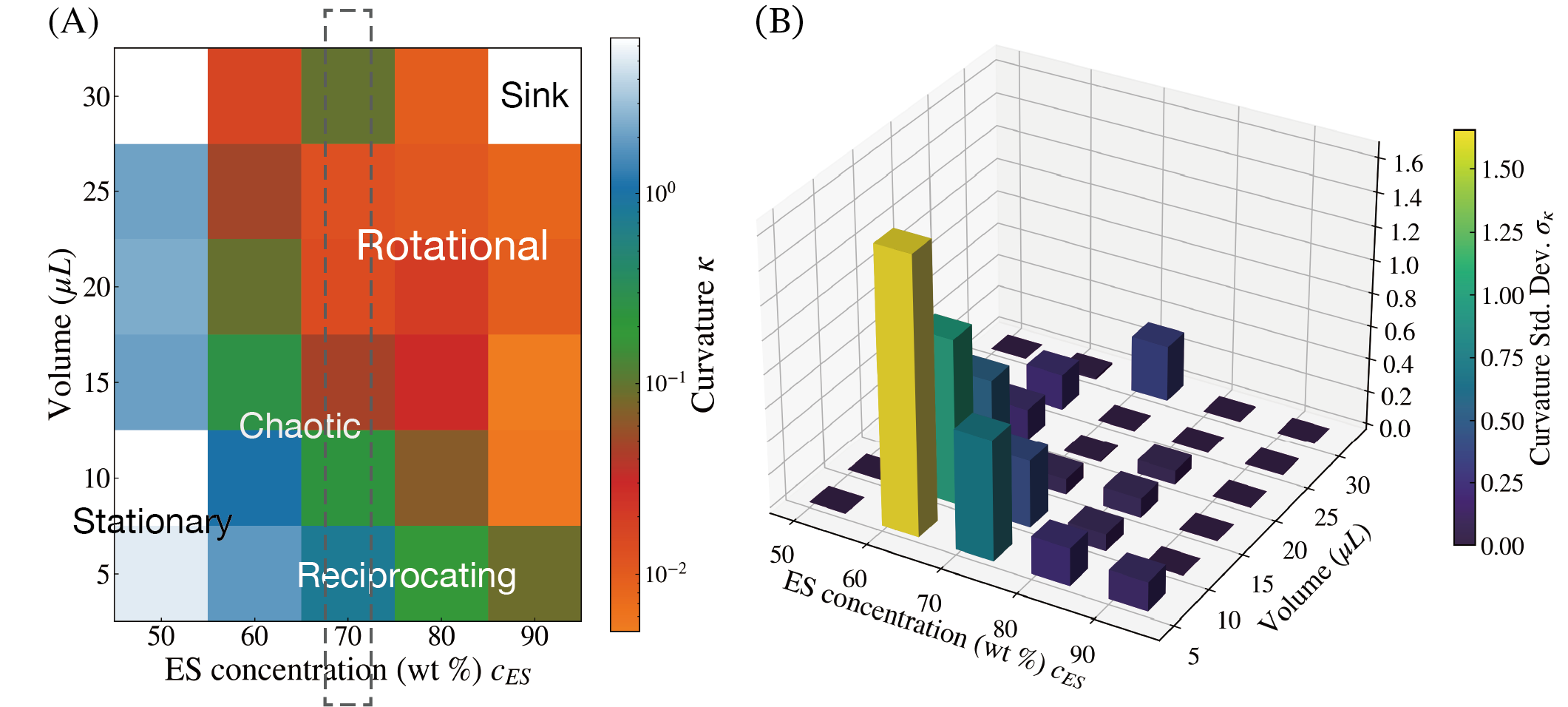}
  \caption{Spatial correspondence to the physical parameter and behavioral patterns. (A) Phase diagram of trajectory patterns depending on experimental configurations. The color code indicates the mean curvature $\langle \kappa \rangle$ over the five trials. Vertical rectangles refer to the cross-section of the phase diagram shown in Fig.\ref{fig:fig7}. (B) 3D bar plot of the standard deviation of the curvature $\sigma_\kappa$ of each experimental configuration.}
  \label{fig:fig6}
\end{figure}

To identify which parameters influence the spontaneous pattern formation of a droplet, we then investigated the responses to the varying droplet volumes and ES concentrations. A phase diagram summarizing the mean curvatures of the trajectories and a resultant class of trajectory patterns obtained after varying configurations of droplet volumes and ES concentrations is presented in Fig.\ref{fig:fig6} (A). Qualitatively consistent patterns emerged near the values shown in each grid, with transitional compositions yielding mixed patterns.

The rotational motions were seen in most parameter sets, namely for large volume ($\ge 20 \mathrm{\, \mu L}$) and/or high ES concentration (80 \%). In the region of medium volumes and ES concentration, the fluctuation of motion becomes evident as quantified by the standard deviations of curvature $\sigma_\kappa$ (Fig.\ref{fig:fig6} (B)), and chaotic patterns emerged. When the volume is small ($\le 10 \mathrm{\, \mu L}$) only, the periodic reciprocation is observed. Once the volume decreases significantly ($\le 5 \mathrm{\, \mu L}$) or the ES concentration drops to 50 \%, the droplet cannot sustain a reaction and enters a stationary state, floating on the surface. Additionally, for volumes over $30 \mathrm{\, \mu L}$ at 90 \% ES concentration, droplets sink under the bulk phase and stay at the same point, corroborating surface tension gradient as the primary driving force.

The oil droplet dynamically modifies the concentration field by dissolving into SDS solutions, and the transition of motion pattern is made unilaterally (e.g., circular to reciprocating or helical to chaotic). Growing the droplet size, enhancing the activity, and reducing the solvent's viscosity resulting from increasing volume and ES concentration might induce directional steady motion. The interplay between droplet volume and solubilization rate dictates whether the motion bifurcates into chaotic or reciprocating patterns. Although not independent, the influence of these manipulable elements on the stability of the motion is shown in Figure \ref{fig:fig6} (B). A continuous chemical reaction sustains the surface tension gradient, but when the surface tension recovery rate matches the reaction rate, the motion becomes intermittent as the surface tension gradient disappears. The droplet's inertia, determined by its volume, influences whether the motion is reciprocating (i.e., autochemotaxis \cite{Tanaka2021-ax}), where the droplet is drawn back to its trajectory, or chaotic with retained periodicity. The transition of motion patterns aligns with previous research investigating developments in the Péclet number \cite{Hokmabad2021-px} with some modification of effective diffusion coefficients \cite{Bickel2019-bs}

\begin{figure}[htbp]
\centering
  \includegraphics[keepaspectratio, width=\linewidth]{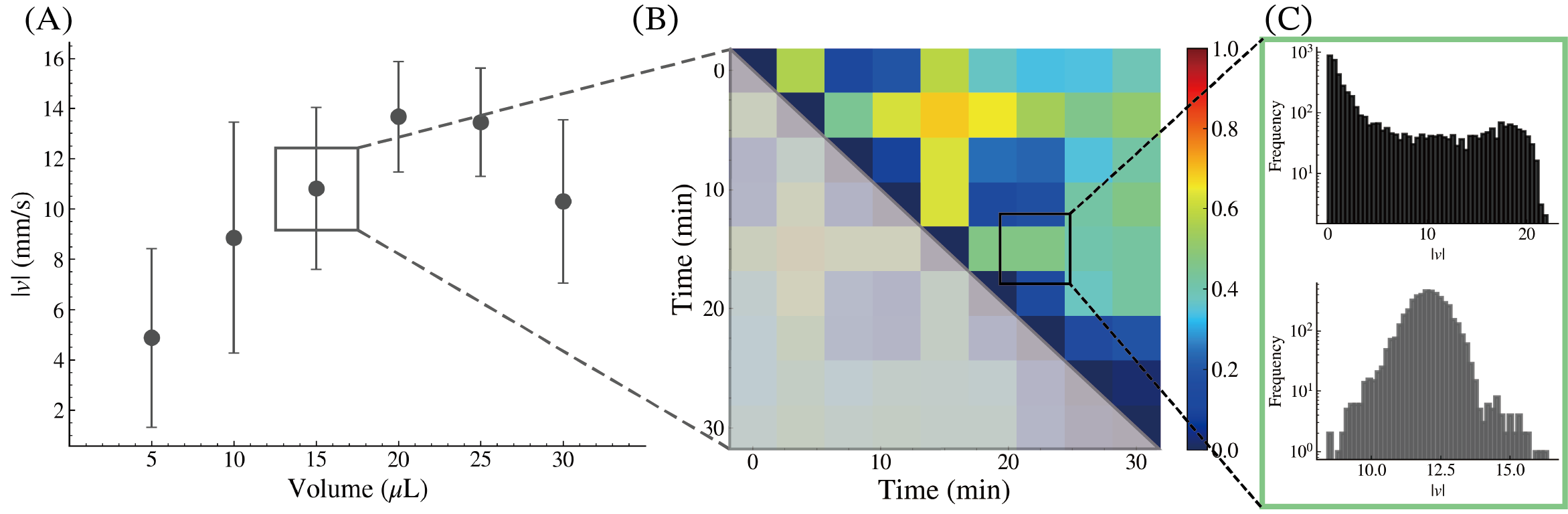}
  \caption{Dependence of physical parameter (volume) on the cruising speed of droplets and its temporal variabilities. (A) Plot of the velocity as a function of the volume. Each data point and error bar illustrate the average and standard deviation derived from 20 to 30 experiments conducted at consistent volumes but varying ES concentrations. (B) Phase diagram of Jensen-Shannon divergence $D_{JS}$ drawn from non-overlapping segments of yielding temporal behavioral variability within an experiment highlighted in a gray box in (A), volume of $15 \mathrm{\, \mu L}$ and an ES concentration of 60 \%. The redundant entries in the lower triangle of the matrix are opaque. (C) Pair of Histograms of velocity for each of the two segments of swimming trajectory in (B) for computation of $D_{JS}$.}
  \label{fig:fig7}
\end{figure}

In Fig.\ref{fig:fig7} (A), we plot the velocity of oil droplets of a given ES concentration against the different volumes (gray dotted box in Fig.\ref{fig:fig6} (A)). We observe the velocity increase rapidly and reach a saturation level at high volume. The evolution of velocity against the ES concentration with a given volume shows a similar tendency (Fig. S1, ESI$\dag$). This result indicates that velocity is the function of the quality and production rate of the droplet left behind the droplet in the aqueous solution. The higher the concentration of reactive agents in the solution, the more stable the enhancement of the surface tension gradient; the larger the droplet, the longer the surface-active substances are removed from the surface. 

The temporal variability of within behavior can be quantified as the distances between two histograms of $|v|$, Jensen-Shannon divergence (JSD) $D_{JS}(P|Q)$ \cite{lin1991divergence}. The JSD evaluates the overlap between two probability distributions (P and Q) and yields 0 for identical distributions and 1 for distributions that do not overlap. By chasing the time course of the JSD, we can obtain insight into the temporal variability of behavioral patterns within an experimental trial. Pairwise comparisons between histograms result in a phase diagram, the one shown in Fig.\ref{fig:fig7} (B), denoted by $D_{JS} (P^N (t) | P^N (t'))$. The structure apparent in the phase diagram in Fig.\ref{fig:fig7} (B) reflects the variation in behavior over the experimental trial of this individual. For example, there is a significant change in the histogram shape due to the transition from circular (Gaussian-like unimodal distribution) to chaotic (bimodal distribution) to circular as shown in Fig.\ref{fig:fig7} (C), the JSD illustrates the change in the behavioral pattern is oriented according to the physical parameters: reductions in either or both the volume and ES concentration drives the droplet susceptible to the external/internal fluctuation, which leads to intermittent motion, eventually shifting the velocity distribution from unimodal to bimodal shape.

\subsection*{Dimensionality of oil droplet behavior}
The above spatiotemporal analysis suggests the deterministic and predictable properties of the "behavioral state-space" of an oil droplet. Here, if we aim to compare with the naturalistic behavior of living organisms, the question arises: how many variables are required to reconstruct the behavioral variability of droplet dynamics? What are the properties of the underlying dynamical systems generating stereotypy and complexity within the various behavioral patterns? To quantitatively explore the underlying dynamical systems of an oil droplet behavior, we employed a data-driven nonparametric approach - empirical dynamic modeling (EDM) \cite{Chang2017-ro}: time-delay embedding \cite{Kennel1992-fd, Sugihara1990-sa} and sequential locally weighted global linear map (S-map) \cite{sugihara1994nonlinear}. We checked all our analysis methods by using them in test data, where the analytic results were known.

\begin{figure}[htbp]
\centering
  \includegraphics[keepaspectratio, width = \linewidth]{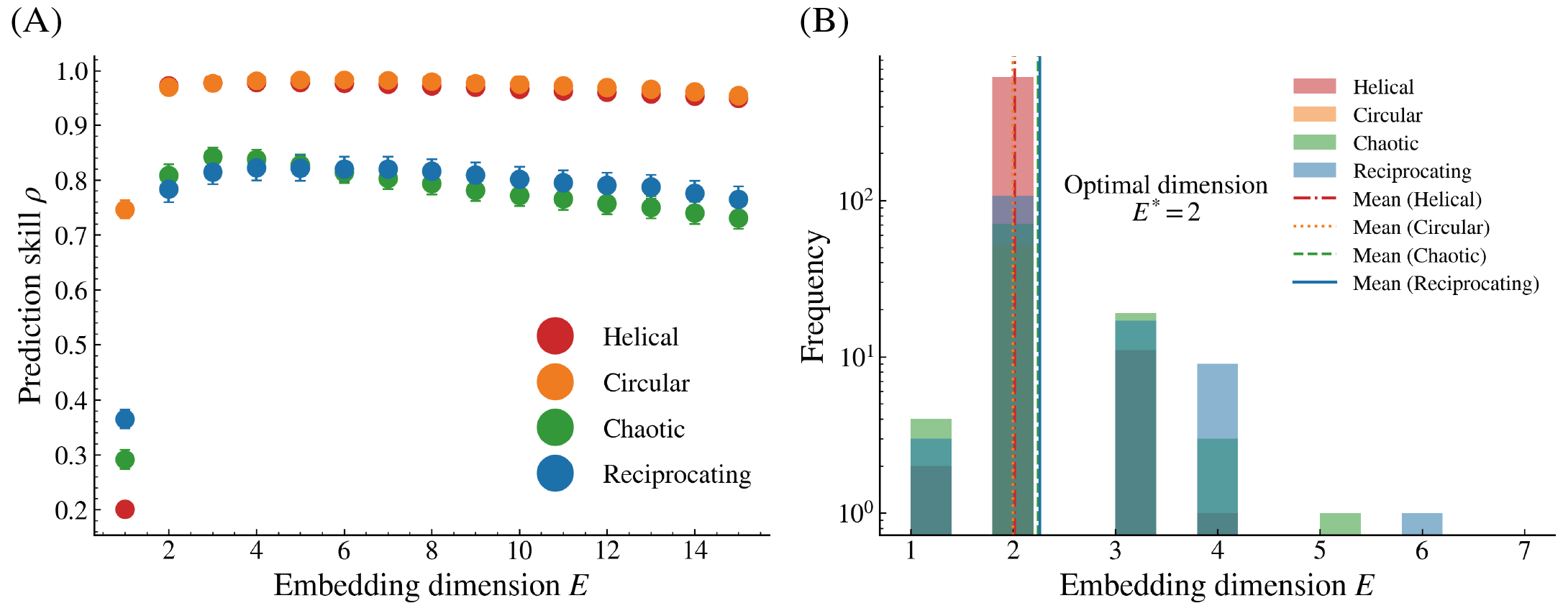}
  \caption{The dimensionality of oil droplet behavior for each trajectory pattern. (A) Estimation of the embedding dimension as the function of prediction skill $\rho$, measured via Pearson correlation coefficient. $\rho$ saturates around the two to three dimensions. Error bars indicate the standard error across all segments. (B) Histogram of the point at which the prediction skill saturates, which we call the optimal embedding dimension $E^{*}$. Periodic rotations exhibit low dimensionality, while chaotic and reciprocating patterns exhibit a statistically significant ($p<0.05$) slightly higher dimensionality $E^{*} \simeq 3$. }
  \label{fig:fig8}
\end{figure}

Fig.\ref{fig:fig8} (A) presents the embedding dimensions of the ensemble average of oil droplet motion within specific behavioral patterns estimated from 1 to 15 dimensions. The time delay $\tau$, an essential parameter in determining the dimensionality, is obtained as the first point where the velocity autocorrelation takes a minimum value (see Fig.\ref{fig:fig11} (G) and Table.\ref{tbl:table1} ACF). The improvement in the prediction skill $\rho$ is saturated at the optimal dimension $E^* = 2$ for almost all cases ($\sim  90 \% $) regardless of trajectory patterns (Fig.\ref{fig:fig8} (B)). This reveals that the oil droplet motion can be characterized within the low dimensional state space. However, some nonlinear complex patterns, including chaotic and reciprocating motions, show $E^{*} = 3$, which is expected based on the Poincaré–Bendixson theorem \cite{strogatz2018nonlinear}. The prediction skills of 80 $\%$ at least indicate that roughly more than 80 $\%$ of the variance in the time series comes from additive noise, which means purely deterministic dynamics. We confirmed that the low dimensionality remains unaffected by the choice of time window $t_w$, and the tendency is consistent within the same parameter configurations, while variations in dimensionality can arise from the fluctuation of the prediction skill, which corresponds to the behavioral mode switching.

\begin{figure*}[htbp]
 \centering
 \includegraphics[width = \linewidth, keepaspectratio]{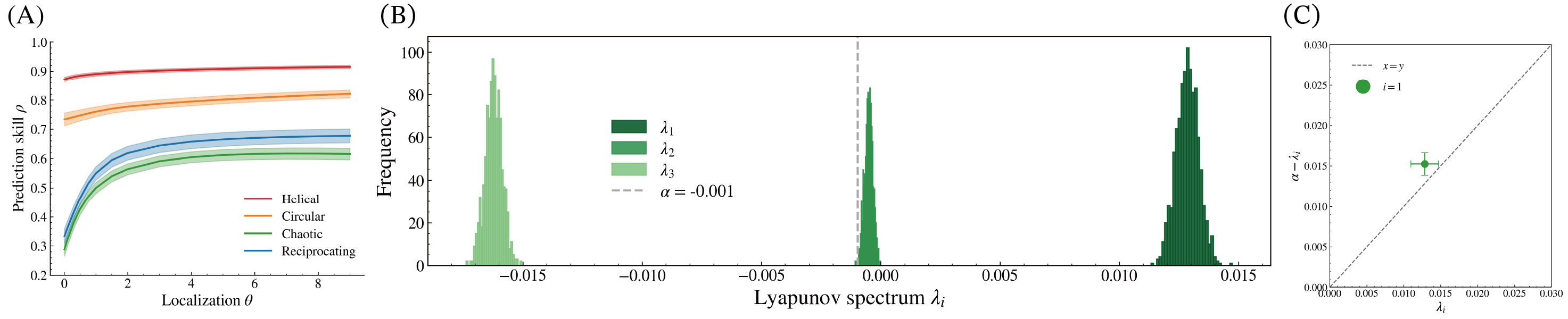}
 \caption{Underlying nonlinear characteristics of dynamical systems of oil droplet behavior. (A) The S-map test for nonlinearity shows the amelioration in prediction skill $\rho$, measured by the correlation coefficient, is observable with increments in the localization parameter $\theta > 0$, which indicative linear equilibrium dynamics for the rotational motion and nonlinear dynamics for the intermittent motion (i.e., chaotic and reciprocating). (B) Lyapunov spectrum of chaotic pattern and its asymmetric structure. Three exponents ($\lambda_i, i = 1,2,3$) were obtained from bootstrap samples, demonstrating near-axis symmetry around $\alpha$ (dashed line). (C) Lyapunov exponents as conjugate pairs that collectively sum up to $\alpha$, akin to a damped-driven Hamiltonian system. Error bars represent 95 \% confidence intervals derived from bootstrapping samples across the entire dataset.}
 \label{fig:fig9}
\end{figure*}

The S-map analysis facilitates quantitatively distinguishing two sources of uncertainty within the system: the inherent nonlinearity of the system itself and the fluctuations associated with its behavior. The S-map analysis facilitates quantitatively distinguishing two sources of uncertainty within the system: the inherent nonlinearity of the system itself and the fluctuations associated with its behavior. S-map is a locally weighted linear regression method that uses all state-space library points with a localization function \cite{sugihara1994nonlinear}:
$$
F(\theta) = \exp(-\theta d/D)
$$
where $d$ is the Euclidean distance, $D$ is the mean distance of all library points, and $\theta$ controls localization. For $\theta = 0$, it reduces to a linear autoregression model, while for $\theta > 0$, predictions depend on the local structure. Comparing fits for $\theta = 0$ (linear) and $\theta > 0$ (nonlinear) reveals the system's state dependency (i.e., nonlinearity) and local state-space structure. Fig.\ref{fig:fig9} (A)  signifies the nonlinearity of chaotic and reciprocating motion and linear characteristics of rotational motion. The prediction skill $\rho$, again measured by the correlation coefficient, improves the localization parameter $\theta$ compared with initial reference $\theta = 0$. This improvement is confirmed as statistically significant ($p<0.05$) compared to the surrogate data fabricated by phase randomization, which retains the original spectrum \cite{prichard1994generating}. The reference range on the reconstructed manifold for the oil droplet dynamics at $\theta = 2$ is limited to a distance of about 20\% compared to the global case ($\theta = 0$), revealing the strong nonlinearity of the system. Nonlinear patterns may originate from different sources of uncertainty depending on inertia from size effect: chaotic patterns arise from chemical reaction inhomogeneities due to increased non-reactive substances, while reciprocating patterns result from weak inertia against the recovery of the surface tension gradient. Key physical parameters, such as ES concentration and volume, play distinct roles in controlling the non-uniformity of chemical reactions. 

\subsection*{Dynamical properties of state space and Lyapunov spectrum}
To investigate the behavioral state-space itself, we computed the spectrum of Lyapunov exponents of the underlying dynamics  \cite{Eckmann1986-wl}. The Lyapunov exponent characterizes a chaotic dynamical system as the separation rate of infinitesimally close trajectories. Positive exponents indicate trajectory bundles expanding along smooth manifolds, while negative ones mean shrinking directions. The interplay between local expansion of trajectory bundles, which creates variability, and overall local contraction due to dissipation, which generates stereotypy, constitutes a fundamental element contributing to the complexity of the swimming dynamics of an oil droplet. The sum of the exponents represents the average dissipation rate, which equals zero in conservative systems and falls below zero in dissipative ones -  symmetry within the Lyapunov spectrum emerges if the conservation law holds (e.g., damped-driven Hamiltonian systems). Fig.\ref{fig:fig9} (B) presents bootstrapped density estimates of the five exponents across all trajectories classified as chaotic. The histograms become opaque as the value of Lyapunov exponents increase $\{ \lambda_i \}_{i =1, 2, 3}$. Among these, one is positive: $\lambda_1 = 0.013$ and one is near-zero exponent: $\lambda_2 = 0.00$. Although we focused on characterizing the three-dimensional embedding, it covers more than 90 \% of the dynamics, our results remain robust and are minimally affected by the choice of the embedding dimension (for four and five dimensions, see Fig. S2, ESI$\dag$). The cumulative value of the positive exponents bounds the Kolmogorov-Sinai (KS) entropy $h_{KS}$, acting as a gauge of unpredictability. The KS entropy was proportional to the increase in volume: $h_{KS} = 0.013$, on average. The summation of all exponents yields a negative value, implying that the system is weakly dissipative, with a dissipation rate of $\sum_{i} \lambda_{i} = -0.004$. The collection of respective values and the spectrum obtained from other behavioral patterns are presented in Fig. S3, ESI$\dag$. In addition, when the dimension of the attractor was evaluated using the Kaplan-York dimension \cite{Frederickson1983-sn}, $D_{KY} = 2.74$ was obtained, which not only coincides with the optimal embedding dimension of the system but is also consistent with the fact that the correlation dimension converges to approximately two dimensions regardless of the behavior pattern.

In Fig.\ref{fig:fig9} (C), the Lyapunov spectrum displays near symmetry, with the pair of exponents closely approaching the boundary of the 95\% bootstrapped confidence interval. These exponents from conjugate pairs sum up to around $\alpha = - 0.001$. The behavioral dynamics of these systems manifest in the breaking of continuous time-reversal symmetry through the dissipation rate $\alpha$. The entire spectrum demonstrates a slight deviation from symmetry, as indicated by the dashed line. Asymmetrical patterns persist across various experimental configurations and trajectory patterns; deviated points correspond to fluctuations in motions, pointing to weak dissipative dynamics. This suggests the unique characteristics of the system as a less dissipative Marangoni surfer among the hydrodynamically most efficient microswimmers. Past studies have reported that the swimming efficiency of the form of a thin circular disk displays the minimum dissipation in terms of Lighthill efficiency \cite{daddi2024hydrodynamic}. In addition, experimental configurations of droplets swimming with high velocity presented in Fig.\ref{fig:fig7} (A) exhibited large deviations from the axis of symmetry and thus imply more dissipative dynamics. We also note that the value of the first Lyapunov exponent takes significantly positive values in the case of chaotic patterns compared with the other periodic motions (Fig. S3, ESI$\dag$), which quantitatively corroborates the chaoticity within the oil droplet behavior. 

Overall, these results revealed that the spatiotemporal variabilities of the oil droplet behavior can be described with the three-dimensional state space. These cover the primary behavioral frequency mode corresponding to the motion stereotypy, and the nonlinearity arises from the fluctuations of chemical reactions manipulated by a droplet's volume and ES concentrations. These physical parameters correspond to the two distinct sources of uncertainty; the lower volume exhibits susceptibility to external gradients, while the inhomogeneity of the internal structure of a droplet internally fluctuates the chemical reaction rate and, thus, unstabilizing the surface tension gradient and interfacial flow around a droplet \cite{Wang2021-jm}. In other words, not only the external stimulus, such as thermal fluctuations, but the self-generating morphology (i.e., the ability to change form in response to the environment to modify dynamics) is fulfilling a critical role in emerging and transitioning droplet motilities. 

\subsection*{Phenomenological description by Langevin equations}
To rationalize the description of the low-dimensionality of oil droplet behavior, we demonstrate the numerical simulation of a phenomenological model of self-propelled droplets. Since our system cannot ignore the inertial effects, we employed the two-dimensional underdamped Langevin equations as a reference point. Taking into account the deviation from the symmetry of the Lyapunov spectrum due to the weak dissipation, we incorporated the driving force into the simplest form of the previous work on Brownian circle swimmer \cite{Van_Teeffelen2008-ya, Wittkowski2012-ed}:
\begin{subequations}
\begin{align}
m\frac{d\mathbf{v}}{dt} &= F \mathbf{\hat{u}} - \gamma \mathbf{v}(t) + \sqrt{2D} \mathbf{\xi}(t) \label{eqn:langevin-1}\\
\frac{d \phi}{dt} &=  M + \sqrt{2D_r} \mathbf{\xi_r}(t) \label{eqn:langevin-2}
\end{align}
\end{subequations}

Where  $m$ is the mass of a droplet, $\mathbf{v} = (v_x, v_y)$  is $x$ and $y$ directional velocity and $\phi$ denotes the orientation of a particle. $\gamma$ is viscosity, $D$ is diffusion coefficient, and $\mathbf{\hat{u}} = (\cos \phi, \sin \phi) $ represents the rotational tensor. $F \mathbf{\hat{u}}$ is a constant driving force responsible for the deterministic motion in the orientation, and $M$ is a constant torque, either internal or external, that leads to rotational motion, which is given as $M = \epsilon |\dot{\phi}|$ for $\epsilon \ll 1$. The fluctuation as non-thermal noise terms $\xi$ and $\xi_r$ were generated as a zero-mean independent Gaussian random number, i.e., $E[\xi(t)] = 0$, $E[\xi(t)\xi(t')] = \delta(t' - t)$. Particle collides elastically according to the law of reflection as a boundary condition. By referring to the experimental observation about the damping coefficient $\alpha$ via the Lyapunov spectrum and for the sake of simplicity, viscosity as a damping factor was fixed to $\gamma = 1$ and $m=1$. The driving forces $|F \mathbf{\hat{u}}|$ are assumed to originate from the constant Marangoni force due to the dissolution and absorption of droplets controlled by volumes and ES concentrations of a droplet. Thus, the parameters can be reduced to the diffusion coefficient $D$ and the amplitude of driving forces $|F| := |F \mathbf{\hat{u}}|$. The Euler-Maruyama method was implemented to integrate a stochastic differential equation numerically.

\begin{figure}[htbp]
\centering
  \includegraphics[keepaspectratio, width=\linewidth]{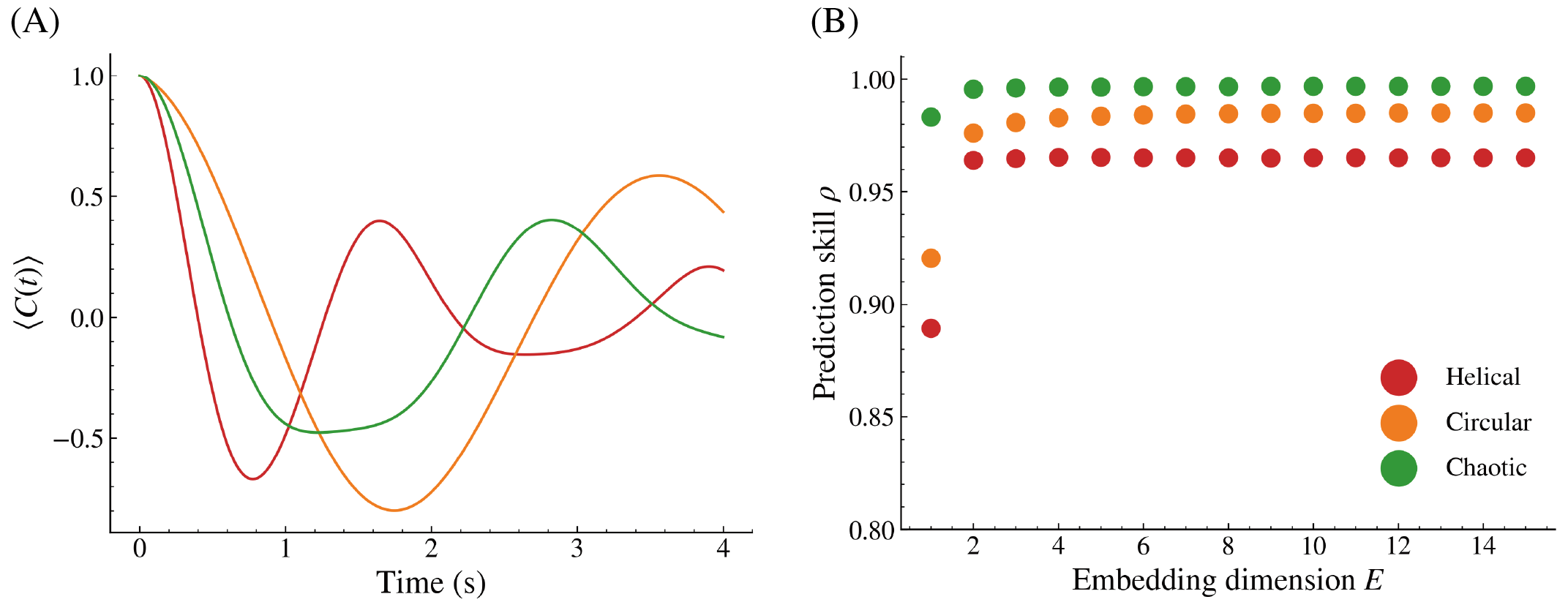}
  \caption{Statistics of simulated two-dimensional trajectories of an oil droplet. (A) Velocity autocorrelation function and (B) embedding dimension estimation of each numerically obtained trajectory pattern. The simulated results and their parameter configurations are presented when circular ($D = 0.03$,  $|F| = 0.08$, and $Pe = 2.7$), helical ($D = 0.09$,  $|F| = 0.1$, and $Pe = 1.1$), and chaotic ($D = 0.03$,  $|F| = 0.06$, and $Pe = 2.0$), respectively.}
  \label{fig:fig10}
\end{figure}

We obtained each trajectory pattern from numerical simulations and calculated the corresponding statistical quantities respectively (Fig.\ref{fig:fig10}). We first reproduced the reciprocating motion by reducing the model to a one-dimensional projection. Subsequently, the two-dimensional rotational motions (i.e., circular and helical) and chaotic patterns are obtained from eqn. (\ref{eqn:langevin-1}, \ref{eqn:langevin-2}) due to the contribution from advection term $|F|$  and superimposed noise amplitudes $D$ (for detailed simulation setting, see Fig.\ref{fig:fig10} and caption). While the simulation results differ from experiments in some aspects, the general phenomenological framework successfully captures linear characteristics of the motion patterns. For instance, chaotic motion characterized by the nonlinearity in the motion (e.g., locally stop-and-go) without contact with a wall of the dish, driven by intrinsic nonlinearity cannot be reproduced by the linear equations used here, as the complexity is instead masked by wall effects. Thus, the autocorrelation function is successfully replicated (Fig.\ref{fig:fig10} (A)); on the other hand, the missing dimensionality in Fig.\ref{fig:fig10} (B), reflects the system's nonlinear characteristics. This highlights the need to incorporate the contributions of internal instabilities, such as non-uniform adsorption and mass transfer, currently excluded from the model in order to capture the full complexity of the system. Rather than fully replicating nonlinear dynamics, the model aims to clarify the fundamental mechanisms underlying a droplet's 'life-like' behavior and identify key nonlinear elements required for future extensions.

From the experimental observation, the behavioral patterns are determined by the balance between inertial forces, influenced by droplet volume, and surface tension gradients, driven by ES concentration. Analogically, assuming the Péclet number $Pe$ is proportional to the ratio of the driving force $|F|$ to diffusion coefficient $D$ (i.e., $Pe := |F|/D$), we can draw the interpretation of correspondence between the simulation parameter and physical ones. Helical patterns appear when both $|F|$ and $D$ are large, which aligns with the fact that they are experimentally observed when the volume and ES concentration are large. Meanwhile, chaotic and circular patterns in the simulation share the same $D= 0.03$ but differ in $|F|$, reflecting differences in droplet volumes. This implies that the circular patterns escape capture by the restored tension gradient due to sufficient inertia. Each pattern indicates a different Péclet number, which is suggestive that there is an intermediate region where inertial effects and chemical reaction instabilities interplay rather than a monotonic increase in motion complexity with $Pe$, leading to chaotic behavior.

Complex behaviors in both nonliving and living systems are often characterized by nonlinearities emerging via interaction with either or both internal/external environments. Here, we showed that the free-running motion of self-propelled oil droplets exhibits nonlinear low-dimensional deterministic properties, as revealed by embedding dimension estimation and S-map analysis. Although droplet's responses to varying physical parameters appear uniform and seemingly dependent on external conditions, transitions between motion patterns can arise from a self-maintaining feedback loop due to internal fluctuations. The coupling of internal morphology with droplet shape deformability may act as an internal state controlling the chemical reaction rate, initiating and sustaining motion. The embedding dimension quantitatively evaluates this behavioral variability, and its fluctuation represents a form of autonomy in a nonliving system, albeit with lower complexity than living systems where both intra and inter-individual variability is crucial. Whereas living systems are expected to continue producing novel behaviors and thus increasing their behavioral variations (also known as open-ended evolution), the limited variability in the embedding dimension of the oil droplets' behavior and the unidirectional transitions in motion patterns may serve as quantitative indicators that distinguish the behavioral dynamics of these systems \cite{Froese2014-ze}. Our phenomenological approach unpacks the limited behavioral discussions of model organisms such as C.elegans into a broader context across scales and species \cite{Stephens2008-et, costa2024markovian}, providing a basis for comparison with the behavioral analysis of living systems. The low-dimensional nature of the droplet's behavior presents a benchmark for developing a physics of naturalistic behavior to differentiate nonliving from living systems quantitatively.





\section{Conclusions}
In this study, we reported the parameter-sweeping experiment and quantitative spatiotemporal characterization of the spontaneous behavior of an oil droplet in an aqueous surfactant solution. We drew a phase diagram by classifying patterned motions, such as rotational (i.e., helical and circular), chaotic, and reciprocating, depending on the physical parameters. The nonlinear dynamical analysis revealed that the underlying dynamical structure of their behavioral patterns is nonlinear and described within low-dimensional state space. The instabilities of internal structure pertaining to the inhomogeneous chemical reaction rate of a droplet can induce chaotic behavior. The nearly symmetric shape of the Lyapunov spectrum is indicative of a unique characteristic of this system driven as a Marangoni surfer. The phenomenological simulation of the two-dimensional Langevin equation captures the linear regime within the dynamics quantitatively. 

Our framework is beneficial for developing future theoretical and experimental research on various spontaneous behavior in general. The natural extension is the collective pattern formation resulting from the environmental modification by mutual interaction between droplets. Also, the system's scalability facilitates scrutinization of the role of droplet large deformation in initiating and sustaining spontaneous motion \cite{Ohta2009-lm, Tarama2016-di}. The long-lasting droplet motion can exhibit non-trivial unique patterns due to the hydrodynamic instabilities of the solubilization, evaporation, and mass transfer. These properties may go beyond a quantitative understanding of the behavior and offer an arena for discussing the essential differences between nonliving matter and living organisms.

\section*{Author contributions}
R.A.: conceptualization, data curation, formal analysis, investigation, methodology, validation, visualization, writing-original draft, writing-review and editing; H.K.: data curation, methodology, supervision, validation, writing-review and editing; T.I.: conceptualization, funding acquisition, project administration, supervision, validation, writing - review and editing. All authors gave final approval for publication and agreed to be held accountable for the work performed therein. 

\section*{Conflicts of interest}
There are no conflicts of interest to declare.

\section*{Data availability}
Presented and analyzed data is included in the main text and the electronic supplementary material. Additional data, including simulations and analysis codes, that support the findings of this study are available from the corresponding author, R.A., upon reasonable request.

\section*{Acknowledgements}
This work was partly supported by Grant-in-Aid for Scientific Research (A) Grant Numbers JP21H04885, JP24H00707. RA was also supported by Grant-in-Aid for JSPS Fellows Grant Number JP24KJ0900. 



\balance

\renewcommand\refname{References}

\bibliography{rsc-articletemplate-softmatter} 

\providecommand*{\mcitethebibliography}{\thebibliography}
\csname @ifundefined\endcsname{endmcitethebibliography}
{\let\endmcitethebibliography\endthebibliography}{}
\begin{mcitethebibliography}{58}
\providecommand*{\natexlab}[1]{#1}
\providecommand*{\mciteSetBstSublistMode}[1]{}
\providecommand*{\mciteSetBstMaxWidthForm}[2]{}
\providecommand*{\mciteBstWouldAddEndPuncttrue}
  {\def\EndOfBibitem{\unskip.}}
\providecommand*{\mciteBstWouldAddEndPunctfalse}
  {\let\EndOfBibitem\relax}
\providecommand*{\mciteSetBstMidEndSepPunct}[3]{}
\providecommand*{\mciteSetBstSublistLabelBeginEnd}[3]{}
\providecommand*{\EndOfBibitem}{}
\mciteSetBstSublistMode{f}
\mciteSetBstMaxWidthForm{subitem}
{(\emph{\alph{mcitesubitemcount}})}
\mciteSetBstSublistLabelBeginEnd{\mcitemaxwidthsubitemform\space}
{\relax}{\relax}

\bibitem[Ramaswamy(2010)]{Ramaswamy2010-ok}
S.~Ramaswamy, \emph{Annual Review of Condensed Matter Physics}, 2010, \textbf{1}, 323--345\relax
\mciteBstWouldAddEndPuncttrue
\mciteSetBstMidEndSepPunct{\mcitedefaultmidpunct}
{\mcitedefaultendpunct}{\mcitedefaultseppunct}\relax
\EndOfBibitem
\bibitem[Shaebani \emph{et~al.}(2020)Shaebani, Wysocki, Winkler, Gompper, and Rieger]{Shaebani2020-et}
M.~R. Shaebani, A.~Wysocki, R.~G. Winkler, G.~Gompper and H.~Rieger, \emph{Nature Reviews Physics}, 2020, \textbf{2}, 181--199\relax
\mciteBstWouldAddEndPuncttrue
\mciteSetBstMidEndSepPunct{\mcitedefaultmidpunct}
{\mcitedefaultendpunct}{\mcitedefaultseppunct}\relax
\EndOfBibitem
\bibitem[Katz \emph{et~al.}(2011)Katz, Tunstrøm, Ioannou, Huepe, and Couzin]{Katz2011-au}
Y.~Katz, K.~Tunstrøm, C.~C. Ioannou, C.~Huepe and I.~D. Couzin, \emph{Proc. Natl. Acad. Sci. U. S. A.}, 2011, \textbf{108}, 18720--18725\relax
\mciteBstWouldAddEndPuncttrue
\mciteSetBstMidEndSepPunct{\mcitedefaultmidpunct}
{\mcitedefaultendpunct}{\mcitedefaultseppunct}\relax
\EndOfBibitem
\bibitem[Ballerini \emph{et~al.}(2008)Ballerini, Cabibbo, Candelier, Cavagna, Cisbani, Giardina, Lecomte, Orlandi, Parisi, Procaccini, Viale, and Zdravkovic]{Ballerini2008-wt}
M.~Ballerini, N.~Cabibbo, R.~Candelier, A.~Cavagna, E.~Cisbani, I.~Giardina, V.~Lecomte, A.~Orlandi, G.~Parisi, A.~Procaccini, M.~Viale and V.~Zdravkovic, \emph{Proc. Natl. Acad. Sci. U. S. A.}, 2008, \textbf{105}, 1232--1237\relax
\mciteBstWouldAddEndPuncttrue
\mciteSetBstMidEndSepPunct{\mcitedefaultmidpunct}
{\mcitedefaultendpunct}{\mcitedefaultseppunct}\relax
\EndOfBibitem
\bibitem[Bowick \emph{et~al.}(2022)Bowick, Fakhri, Marchetti, and Ramaswamy]{Bowick2022-ti}
M.~J. Bowick, N.~Fakhri, M.~C. Marchetti and S.~Ramaswamy, \emph{Phys. Rev. X}, 2022, \textbf{12}, 010501\relax
\mciteBstWouldAddEndPuncttrue
\mciteSetBstMidEndSepPunct{\mcitedefaultmidpunct}
{\mcitedefaultendpunct}{\mcitedefaultseppunct}\relax
\EndOfBibitem
\bibitem[Marchetti \emph{et~al.}(2013)Marchetti, Joanny, Ramaswamy, Liverpool, Prost, Rao, and Simha]{Marchetti2013-da}
M.~C. Marchetti, J.~F. Joanny, S.~Ramaswamy, T.~B. Liverpool, J.~Prost, M.~Rao and R.~A. Simha, \emph{Rev. Mod. Phys.}, 2013, \textbf{85}, 1143--1189\relax
\mciteBstWouldAddEndPuncttrue
\mciteSetBstMidEndSepPunct{\mcitedefaultmidpunct}
{\mcitedefaultendpunct}{\mcitedefaultseppunct}\relax
\EndOfBibitem
\bibitem[Tu and Rappel(2018)]{Tu2018-vu}
Y.~Tu and W.-J. Rappel, \emph{Annu Rev Condens Matter Phys}, 2018, \textbf{9}, 183--205\relax
\mciteBstWouldAddEndPuncttrue
\mciteSetBstMidEndSepPunct{\mcitedefaultmidpunct}
{\mcitedefaultendpunct}{\mcitedefaultseppunct}\relax
\EndOfBibitem
\bibitem[Maass \emph{et~al.}(2016)Maass, Krüger, Herminghaus, and Bahr]{Maass2016-wl}
C.~C. Maass, C.~Krüger, S.~Herminghaus and C.~Bahr, \emph{Annual Review of Condensed Matter Physics}, 2016, \textbf{7}, 171--193\relax
\mciteBstWouldAddEndPuncttrue
\mciteSetBstMidEndSepPunct{\mcitedefaultmidpunct}
{\mcitedefaultendpunct}{\mcitedefaultseppunct}\relax
\EndOfBibitem
\bibitem[Hokmabad \emph{et~al.}(2021)Hokmabad, Dey, Jalaal, Mohanty, Almukambetova, Baldwin, Lohse, and Maass]{Hokmabad2021-px}
B.~V. Hokmabad, R.~Dey, M.~Jalaal, D.~Mohanty, M.~Almukambetova, K.~A. Baldwin, D.~Lohse and C.~C. Maass, \emph{Phys. Rev. X}, 2021, \textbf{11}, 011043\relax
\mciteBstWouldAddEndPuncttrue
\mciteSetBstMidEndSepPunct{\mcitedefaultmidpunct}
{\mcitedefaultendpunct}{\mcitedefaultseppunct}\relax
\EndOfBibitem
\bibitem[Michelin(2023)]{Michelin2023-ut}
S.~Michelin, \emph{Annu. Rev. Fluid Mech.}, 2023, \textbf{55}, 77--101\relax
\mciteBstWouldAddEndPuncttrue
\mciteSetBstMidEndSepPunct{\mcitedefaultmidpunct}
{\mcitedefaultendpunct}{\mcitedefaultseppunct}\relax
\EndOfBibitem
\bibitem[Toyota \emph{et~al.}(2009)Toyota, Maru, Hanczyc, Ikegami, and Sugawara]{Toyota2009-ge}
T.~Toyota, N.~Maru, M.~M. Hanczyc, T.~Ikegami and T.~Sugawara, \emph{J. Am. Chem. Soc.}, 2009, \textbf{131}, 5012--5013\relax
\mciteBstWouldAddEndPuncttrue
\mciteSetBstMidEndSepPunct{\mcitedefaultmidpunct}
{\mcitedefaultendpunct}{\mcitedefaultseppunct}\relax
\EndOfBibitem
\bibitem[Horibe \emph{et~al.}(2011)Horibe, Hanczyc, and Ikegami]{Horibe2011-sy}
N.~Horibe, M.~M. Hanczyc and T.~Ikegami, \emph{Entropy}, 2011, \textbf{13}, 709--719\relax
\mciteBstWouldAddEndPuncttrue
\mciteSetBstMidEndSepPunct{\mcitedefaultmidpunct}
{\mcitedefaultendpunct}{\mcitedefaultseppunct}\relax
\EndOfBibitem
\bibitem[Sumino \emph{et~al.}(2005)Sumino, Magome, Hamada, and Yoshikawa]{Sumino2005-ep}
Y.~Sumino, N.~Magome, T.~Hamada and K.~Yoshikawa, \emph{Phys. Rev. Lett.}, 2005, \textbf{94}, 068301\relax
\mciteBstWouldAddEndPuncttrue
\mciteSetBstMidEndSepPunct{\mcitedefaultmidpunct}
{\mcitedefaultendpunct}{\mcitedefaultseppunct}\relax
\EndOfBibitem
\bibitem[Nagai \emph{et~al.}(2005)Nagai, Sumino, Kitahata, and Yoshikawa]{Nagai2005-zy}
K.~Nagai, Y.~Sumino, H.~Kitahata and K.~Yoshikawa, \emph{Phys. Rev. E Stat. Nonlin. Soft Matter Phys.}, 2005, \textbf{71}, 065301\relax
\mciteBstWouldAddEndPuncttrue
\mciteSetBstMidEndSepPunct{\mcitedefaultmidpunct}
{\mcitedefaultendpunct}{\mcitedefaultseppunct}\relax
\EndOfBibitem
\bibitem[Krüger \emph{et~al.}(2016)Krüger, Bahr, Herminghaus, and Maass]{Kruger2016-lq}
C.~Krüger, C.~Bahr, S.~Herminghaus and C.~C. Maass, \emph{Eur. Phys. J. E Soft Matter}, 2016, \textbf{39}, 64\relax
\mciteBstWouldAddEndPuncttrue
\mciteSetBstMidEndSepPunct{\mcitedefaultmidpunct}
{\mcitedefaultendpunct}{\mcitedefaultseppunct}\relax
\EndOfBibitem
\bibitem[Suda \emph{et~al.}(2022)Suda, Suda, Ohmura, and Ichikawa]{Suda2022-yz}
S.~Suda, T.~Suda, T.~Ohmura and M.~Ichikawa, \emph{Phys Rev E}, 2022, \textbf{106}, 034610\relax
\mciteBstWouldAddEndPuncttrue
\mciteSetBstMidEndSepPunct{\mcitedefaultmidpunct}
{\mcitedefaultendpunct}{\mcitedefaultseppunct}\relax
\EndOfBibitem
\bibitem[Hanczyc \emph{et~al.}(2007)Hanczyc, Toyota, Ikegami, Packard, and Sugawara]{Hanczyc2007-pc}
M.~M. Hanczyc, T.~Toyota, T.~Ikegami, N.~Packard and T.~Sugawara, \emph{J. Am. Chem. Soc.}, 2007, \textbf{129}, 9386--9391\relax
\mciteBstWouldAddEndPuncttrue
\mciteSetBstMidEndSepPunct{\mcitedefaultmidpunct}
{\mcitedefaultendpunct}{\mcitedefaultseppunct}\relax
\EndOfBibitem
\bibitem[Izri \emph{et~al.}(2014)Izri, van~der Linden, Michelin, and Dauchot]{Izri2014-vc}
Z.~Izri, M.~N. van~der Linden, S.~Michelin and O.~Dauchot, \emph{Phys. Rev. Lett.}, 2014, \textbf{113}, 248302\relax
\mciteBstWouldAddEndPuncttrue
\mciteSetBstMidEndSepPunct{\mcitedefaultmidpunct}
{\mcitedefaultendpunct}{\mcitedefaultseppunct}\relax
\EndOfBibitem
\bibitem[Hanczyc \emph{et~al.}(2003)Hanczyc, Fujikawa, and Szostak]{Hanczyc2003-ra}
M.~Hanczyc, S.~M. Fujikawa and J.~Szostak, \emph{Science}, 2003, \textbf{302}, 618--622\relax
\mciteBstWouldAddEndPuncttrue
\mciteSetBstMidEndSepPunct{\mcitedefaultmidpunct}
{\mcitedefaultendpunct}{\mcitedefaultseppunct}\relax
\EndOfBibitem
\bibitem[Zwicker \emph{et~al.}(2016)Zwicker, Seyboldt, Weber, Hyman, and Jülicher]{Zwicker2016-ic}
D.~Zwicker, R.~Seyboldt, C.~A. Weber, A.~A. Hyman and F.~Jülicher, \emph{Nat. Phys.}, 2016, \textbf{13}, 408--413\relax
\mciteBstWouldAddEndPuncttrue
\mciteSetBstMidEndSepPunct{\mcitedefaultmidpunct}
{\mcitedefaultendpunct}{\mcitedefaultseppunct}\relax
\EndOfBibitem
\bibitem[Meredith \emph{et~al.}(2020)Meredith, Moerman, Groenewold, Chiu, Kegel, van Blaaderen, and Zarzar]{Meredith2020-ur}
C.~H. Meredith, P.~G. Moerman, J.~Groenewold, Y.-J. Chiu, W.~K. Kegel, A.~van Blaaderen and L.~D. Zarzar, \emph{Nat. Chem.}, 2020, \textbf{12}, 1136--1142\relax
\mciteBstWouldAddEndPuncttrue
\mciteSetBstMidEndSepPunct{\mcitedefaultmidpunct}
{\mcitedefaultendpunct}{\mcitedefaultseppunct}\relax
\EndOfBibitem
\bibitem[Liu \emph{et~al.}(2024)Liu, Kailasham, Moerman, Khair, and Zarzar]{Liu2024-kt}
Y.~Liu, R.~Kailasham, P.~G. Moerman, A.~S. Khair and L.~Zarzar, \emph{Angew. Chem. Int. Ed Engl.}, 2024,  e202409382\relax
\mciteBstWouldAddEndPuncttrue
\mciteSetBstMidEndSepPunct{\mcitedefaultmidpunct}
{\mcitedefaultendpunct}{\mcitedefaultseppunct}\relax
\EndOfBibitem
\bibitem[Lagzi \emph{et~al.}(2010)Lagzi, Soh, Wesson, Browne, and Grzybowski]{Lagzi2010-ad}
I.~Lagzi, S.~Soh, P.~J. Wesson, K.~P. Browne and B.~A. Grzybowski, \emph{J. Am. Chem. Soc.}, 2010, \textbf{132}, 1198--1199\relax
\mciteBstWouldAddEndPuncttrue
\mciteSetBstMidEndSepPunct{\mcitedefaultmidpunct}
{\mcitedefaultendpunct}{\mcitedefaultseppunct}\relax
\EndOfBibitem
\bibitem[Jin \emph{et~al.}(2017)Jin, Krüger, and Maass]{Jin2017-yn}
C.~Jin, C.~Krüger and C.~C. Maass, \emph{Proc. Natl. Acad. Sci. U. S. A.}, 2017, \textbf{114}, 5089--5094\relax
\mciteBstWouldAddEndPuncttrue
\mciteSetBstMidEndSepPunct{\mcitedefaultmidpunct}
{\mcitedefaultendpunct}{\mcitedefaultseppunct}\relax
\EndOfBibitem
\bibitem[Tanaka \emph{et~al.}(2017)Tanaka, Nakata, and Kano]{Tanaka2017-vu}
S.~Tanaka, S.~Nakata and T.~Kano, \emph{J. Phys. Soc. Jpn.}, 2017, \textbf{86}, 101004\relax
\mciteBstWouldAddEndPuncttrue
\mciteSetBstMidEndSepPunct{\mcitedefaultmidpunct}
{\mcitedefaultendpunct}{\mcitedefaultseppunct}\relax
\EndOfBibitem
\bibitem[Jin \emph{et~al.}(2021)Jin, Chen, Maass, and Mathijssen]{Jin2021-fn}
C.~Jin, Y.~Chen, C.~C. Maass and A.~J. T.~M. Mathijssen, \emph{Phys. Rev. Lett.}, 2021, \textbf{127}, 088006\relax
\mciteBstWouldAddEndPuncttrue
\mciteSetBstMidEndSepPunct{\mcitedefaultmidpunct}
{\mcitedefaultendpunct}{\mcitedefaultseppunct}\relax
\EndOfBibitem
\bibitem[Pimienta and Antoine(2014)]{Pimienta2014-fh}
V.~Pimienta and C.~Antoine, \emph{Curr. Opin. Colloid Interface Sci.}, 2014, \textbf{19}, 290--299\relax
\mciteBstWouldAddEndPuncttrue
\mciteSetBstMidEndSepPunct{\mcitedefaultmidpunct}
{\mcitedefaultendpunct}{\mcitedefaultseppunct}\relax
\EndOfBibitem
\bibitem[Hanczyc and Ikegami(2010)]{Hanczyc2010-jx}
M.~M. Hanczyc and T.~Ikegami, \emph{Artif. Life}, 2010, \textbf{16}, 233--243\relax
\mciteBstWouldAddEndPuncttrue
\mciteSetBstMidEndSepPunct{\mcitedefaultmidpunct}
{\mcitedefaultendpunct}{\mcitedefaultseppunct}\relax
\EndOfBibitem
\bibitem[Suematsu and Nakata(2018)]{suematsu2018evolution}
N.~J. Suematsu and S.~Nakata, \emph{Chemistry--A European Journal}, 2018, \textbf{24}, 6308--6324\relax
\mciteBstWouldAddEndPuncttrue
\mciteSetBstMidEndSepPunct{\mcitedefaultmidpunct}
{\mcitedefaultendpunct}{\mcitedefaultseppunct}\relax
\EndOfBibitem
\bibitem[Berman \emph{et~al.}(2016)Berman, Bialek, and Shaevitz]{Berman2016-ut}
G.~J. Berman, W.~Bialek and J.~W. Shaevitz, \emph{Proc. Natl. Acad. Sci. U. S. A.}, 2016, \textbf{113}, 11943--11948\relax
\mciteBstWouldAddEndPuncttrue
\mciteSetBstMidEndSepPunct{\mcitedefaultmidpunct}
{\mcitedefaultendpunct}{\mcitedefaultseppunct}\relax
\EndOfBibitem
\bibitem[Bialek(2022)]{Bialek2022-pl}
W.~Bialek, \emph{Proc. Natl. Acad. Sci. U. S. A.}, 2022, \textbf{119}, e2021860119\relax
\mciteBstWouldAddEndPuncttrue
\mciteSetBstMidEndSepPunct{\mcitedefaultmidpunct}
{\mcitedefaultendpunct}{\mcitedefaultseppunct}\relax
\EndOfBibitem
\bibitem[Ahamed \emph{et~al.}(2020)Ahamed, Costa, and Stephens]{Ahamed2020-np}
T.~Ahamed, A.~C. Costa and G.~J. Stephens, \emph{Nat. Phys.}, 2020, \textbf{17}, 275--283\relax
\mciteBstWouldAddEndPuncttrue
\mciteSetBstMidEndSepPunct{\mcitedefaultmidpunct}
{\mcitedefaultendpunct}{\mcitedefaultseppunct}\relax
\EndOfBibitem
\bibitem[Berman \emph{et~al.}(2014)Berman, Choi, Bialek, and Shaevitz]{berman2014mapping}
G.~J. Berman, D.~M. Choi, W.~Bialek and J.~W. Shaevitz, \emph{Journal of The Royal Society Interface}, 2014, \textbf{11}, 20140672\relax
\mciteBstWouldAddEndPuncttrue
\mciteSetBstMidEndSepPunct{\mcitedefaultmidpunct}
{\mcitedefaultendpunct}{\mcitedefaultseppunct}\relax
\EndOfBibitem
\bibitem[Tanaka \emph{et~al.}(2015)Tanaka, Sogabe, and Nakata]{Tanaka2015-sr}
S.~Tanaka, Y.~Sogabe and S.~Nakata, \emph{Phys. Rev. E Stat. Nonlin. Soft Matter Phys.}, 2015, \textbf{91}, 032406\relax
\mciteBstWouldAddEndPuncttrue
\mciteSetBstMidEndSepPunct{\mcitedefaultmidpunct}
{\mcitedefaultendpunct}{\mcitedefaultseppunct}\relax
\EndOfBibitem
\bibitem[Satoh \emph{et~al.}(2017)Satoh, Sogabe, Kayahara, Tanaka, Nagayama, and Nakata]{Satoh2017-oz}
Y.~Satoh, Y.~Sogabe, K.~Kayahara, S.~Tanaka, M.~Nagayama and S.~Nakata, \emph{Soft Matter}, 2017, \textbf{13}, 3422--3430\relax
\mciteBstWouldAddEndPuncttrue
\mciteSetBstMidEndSepPunct{\mcitedefaultmidpunct}
{\mcitedefaultendpunct}{\mcitedefaultseppunct}\relax
\EndOfBibitem
\bibitem[Tanaka \emph{et~al.}(2021)Tanaka, Nakata, and Nagayama]{Tanaka2021-ax}
S.~Tanaka, S.~Nakata and M.~Nagayama, \emph{Soft Matter}, 2021, \textbf{17}, 388--396\relax
\mciteBstWouldAddEndPuncttrue
\mciteSetBstMidEndSepPunct{\mcitedefaultmidpunct}
{\mcitedefaultendpunct}{\mcitedefaultseppunct}\relax
\EndOfBibitem
\bibitem[Marumo \emph{et~al.}(2021)Marumo, Yamagishi, and Yajima]{Marumo2021-ua}
A.~Marumo, M.~Yamagishi and J.~Yajima, \emph{Commun Biol}, 2021, \textbf{4}, 1209\relax
\mciteBstWouldAddEndPuncttrue
\mciteSetBstMidEndSepPunct{\mcitedefaultmidpunct}
{\mcitedefaultendpunct}{\mcitedefaultseppunct}\relax
\EndOfBibitem
\bibitem[Lauga and Powers(2009)]{Lauga2009-iz}
E.~Lauga and T.~R. Powers, \emph{Rep. Prog. Phys.}, 2009, \textbf{72}, 096601\relax
\mciteBstWouldAddEndPuncttrue
\mciteSetBstMidEndSepPunct{\mcitedefaultmidpunct}
{\mcitedefaultendpunct}{\mcitedefaultseppunct}\relax
\EndOfBibitem
\bibitem[Perez~Ipiña \emph{et~al.}(2019)Perez~Ipiña, Otte, Pontier-Bres, Czerucka, and Peruani]{Perez_Ipina2019-ot}
E.~Perez~Ipiña, S.~Otte, R.~Pontier-Bres, D.~Czerucka and F.~Peruani, \emph{Nat. Phys.}, 2019, \textbf{15}, 610--615\relax
\mciteBstWouldAddEndPuncttrue
\mciteSetBstMidEndSepPunct{\mcitedefaultmidpunct}
{\mcitedefaultendpunct}{\mcitedefaultseppunct}\relax
\EndOfBibitem
\bibitem[Bickel(2019)]{Bickel2019-bs}
T.~Bickel, \emph{Soft Matter}, 2019, \textbf{15}, 3644--3648\relax
\mciteBstWouldAddEndPuncttrue
\mciteSetBstMidEndSepPunct{\mcitedefaultmidpunct}
{\mcitedefaultendpunct}{\mcitedefaultseppunct}\relax
\EndOfBibitem
\bibitem[Lin(1991)]{lin1991divergence}
J.~Lin, \emph{IEEE Transactions on Information theory}, 1991, \textbf{37}, 145--151\relax
\mciteBstWouldAddEndPuncttrue
\mciteSetBstMidEndSepPunct{\mcitedefaultmidpunct}
{\mcitedefaultendpunct}{\mcitedefaultseppunct}\relax
\EndOfBibitem
\bibitem[Chang \emph{et~al.}(2017)Chang, Ushio, and Hsieh]{Chang2017-ro}
C.-W. Chang, M.~Ushio and C.-H. Hsieh, \emph{Ecol. Res.}, 2017, \textbf{32}, 785--796\relax
\mciteBstWouldAddEndPuncttrue
\mciteSetBstMidEndSepPunct{\mcitedefaultmidpunct}
{\mcitedefaultendpunct}{\mcitedefaultseppunct}\relax
\EndOfBibitem
\bibitem[Kennel \emph{et~al.}(1992)Kennel, Brown, and Abarbanel]{Kennel1992-fd}
M.~B. Kennel, R.~Brown and H.~D. Abarbanel, \emph{Phys. Rev. A}, 1992, \textbf{45}, 3403--3411\relax
\mciteBstWouldAddEndPuncttrue
\mciteSetBstMidEndSepPunct{\mcitedefaultmidpunct}
{\mcitedefaultendpunct}{\mcitedefaultseppunct}\relax
\EndOfBibitem
\bibitem[Sugihara and May(1990)]{Sugihara1990-sa}
G.~Sugihara and R.~M. May, \emph{Nature}, 1990, \textbf{344}, 734--741\relax
\mciteBstWouldAddEndPuncttrue
\mciteSetBstMidEndSepPunct{\mcitedefaultmidpunct}
{\mcitedefaultendpunct}{\mcitedefaultseppunct}\relax
\EndOfBibitem
\bibitem[Sugihara(1994)]{sugihara1994nonlinear}
G.~Sugihara, \emph{Philosophical Transactions of the Royal Society of London. Series A: Physical and Engineering Sciences}, 1994, \textbf{348}, 477--495\relax
\mciteBstWouldAddEndPuncttrue
\mciteSetBstMidEndSepPunct{\mcitedefaultmidpunct}
{\mcitedefaultendpunct}{\mcitedefaultseppunct}\relax
\EndOfBibitem
\bibitem[Strogatz(2018)]{strogatz2018nonlinear}
S.~H. Strogatz, \emph{Nonlinear dynamics and chaos: with applications to physics, biology, chemistry, and engineering}, CRC press, 2018\relax
\mciteBstWouldAddEndPuncttrue
\mciteSetBstMidEndSepPunct{\mcitedefaultmidpunct}
{\mcitedefaultendpunct}{\mcitedefaultseppunct}\relax
\EndOfBibitem
\bibitem[Prichard and Theiler(1994)]{prichard1994generating}
D.~Prichard and J.~Theiler, \emph{Physical review letters}, 1994, \textbf{73}, 951\relax
\mciteBstWouldAddEndPuncttrue
\mciteSetBstMidEndSepPunct{\mcitedefaultmidpunct}
{\mcitedefaultendpunct}{\mcitedefaultseppunct}\relax
\EndOfBibitem
\bibitem[Eckmann \emph{et~al.}(1986)Eckmann, Kamphorst, Ruelle, and Ciliberto]{Eckmann1986-wl}
J.~Eckmann, S.~O. Kamphorst, D.~Ruelle and S.~Ciliberto, \emph{Phys. Rev. A Gen. Phys.}, 1986, \textbf{34}, 4971--4979\relax
\mciteBstWouldAddEndPuncttrue
\mciteSetBstMidEndSepPunct{\mcitedefaultmidpunct}
{\mcitedefaultendpunct}{\mcitedefaultseppunct}\relax
\EndOfBibitem
\bibitem[Frederickson \emph{et~al.}(1983)Frederickson, Kaplan, Yorke, and Yorke]{Frederickson1983-sn}
P.~Frederickson, J.~Kaplan, E.~Yorke and J.~Yorke, \emph{J. Differ. Equ.}, 1983, \textbf{49}, 185--207\relax
\mciteBstWouldAddEndPuncttrue
\mciteSetBstMidEndSepPunct{\mcitedefaultmidpunct}
{\mcitedefaultendpunct}{\mcitedefaultseppunct}\relax
\EndOfBibitem
\bibitem[Daddi-Moussa-Ider \emph{et~al.}(2024)Daddi-Moussa-Ider, Golestanian, and Vilfan]{daddi2024hydrodynamic}
A.~Daddi-Moussa-Ider, R.~Golestanian and A.~Vilfan, \emph{Journal of Fluid Mechanics}, 2024, \textbf{986}, A32\relax
\mciteBstWouldAddEndPuncttrue
\mciteSetBstMidEndSepPunct{\mcitedefaultmidpunct}
{\mcitedefaultendpunct}{\mcitedefaultseppunct}\relax
\EndOfBibitem
\bibitem[Wang \emph{et~al.}(2021)Wang, Zhang, Mozaffari, de~Pablo, and Abbott]{Wang2021-jm}
X.~Wang, R.~Zhang, A.~Mozaffari, J.~J. de~Pablo and N.~L. Abbott, \emph{Soft Matter}, 2021, \textbf{17}, 2985--2993\relax
\mciteBstWouldAddEndPuncttrue
\mciteSetBstMidEndSepPunct{\mcitedefaultmidpunct}
{\mcitedefaultendpunct}{\mcitedefaultseppunct}\relax
\EndOfBibitem
\bibitem[van Teeffelen and Löwen(2008)]{Van_Teeffelen2008-ya}
S.~van Teeffelen and H.~Löwen, \emph{Phys. Rev. E Stat. Nonlin. Soft Matter Phys.}, 2008, \textbf{78}, 020101\relax
\mciteBstWouldAddEndPuncttrue
\mciteSetBstMidEndSepPunct{\mcitedefaultmidpunct}
{\mcitedefaultendpunct}{\mcitedefaultseppunct}\relax
\EndOfBibitem
\bibitem[Wittkowski and Löwen(2012)]{Wittkowski2012-ed}
R.~Wittkowski and H.~Löwen, \emph{Phys. Rev. E Stat. Nonlin. Soft Matter Phys.}, 2012, \textbf{85}, 021406\relax
\mciteBstWouldAddEndPuncttrue
\mciteSetBstMidEndSepPunct{\mcitedefaultmidpunct}
{\mcitedefaultendpunct}{\mcitedefaultseppunct}\relax
\EndOfBibitem
\bibitem[Froese \emph{et~al.}(2014)Froese, Virgo, and Ikegami]{Froese2014-ze}
T.~Froese, N.~Virgo and T.~Ikegami, \emph{Artif. Life}, 2014, \textbf{20}, 55--76\relax
\mciteBstWouldAddEndPuncttrue
\mciteSetBstMidEndSepPunct{\mcitedefaultmidpunct}
{\mcitedefaultendpunct}{\mcitedefaultseppunct}\relax
\EndOfBibitem
\bibitem[Stephens \emph{et~al.}(2008)Stephens, Johnson-Kerner, Bialek, and Ryu]{Stephens2008-et}
G.~J. Stephens, B.~Johnson-Kerner, W.~Bialek and W.~S. Ryu, \emph{PLoS Comput. Biol.}, 2008, \textbf{4}, e1000028\relax
\mciteBstWouldAddEndPuncttrue
\mciteSetBstMidEndSepPunct{\mcitedefaultmidpunct}
{\mcitedefaultendpunct}{\mcitedefaultseppunct}\relax
\EndOfBibitem
\bibitem[Costa \emph{et~al.}(2024)Costa, Ahamed, Jordan, and Stephens]{costa2024markovian}
A.~C. Costa, T.~Ahamed, D.~Jordan and G.~J. Stephens, \emph{Proceedings of the National Academy of Sciences}, 2024, \textbf{121}, e2318805121\relax
\mciteBstWouldAddEndPuncttrue
\mciteSetBstMidEndSepPunct{\mcitedefaultmidpunct}
{\mcitedefaultendpunct}{\mcitedefaultseppunct}\relax
\EndOfBibitem
\bibitem[Ohta and Ohkuma(2009)]{Ohta2009-lm}
T.~Ohta and T.~Ohkuma, \emph{Phys. Rev. Lett.}, 2009, \textbf{102}, 154101\relax
\mciteBstWouldAddEndPuncttrue
\mciteSetBstMidEndSepPunct{\mcitedefaultmidpunct}
{\mcitedefaultendpunct}{\mcitedefaultseppunct}\relax
\EndOfBibitem
\bibitem[Tarama and Ohta(2016)]{Tarama2016-di}
M.~Tarama and T.~Ohta, \emph{EPL}, 2016, \textbf{114}, 30002\relax
\mciteBstWouldAddEndPuncttrue
\mciteSetBstMidEndSepPunct{\mcitedefaultmidpunct}
{\mcitedefaultendpunct}{\mcitedefaultseppunct}\relax
\EndOfBibitem
\end{mcitethebibliography}
\bibliographystyle{rsc} 

\end{document}